 \documentclass[final,authoryear,3p,times,twocolumn]{elsarticle}

\usepackage{graphicx}

\usepackage{amssymb}

\usepackage{rotating}
\usepackage{array}
\usepackage{color}
\usepackage{multirow}
\usepackage{float}
\usepackage{url}
\usepackage{ulem}
\usepackage{subfigure}
\usepackage{hyperref}
\hypersetup{
bookmarks=true,        
unicode=false,          
pdftoolbar=true,        
pdfmenubar=true,        
pdffitwindow=false,     
pdfstartview={FitH},    
pdftitle={N2 emission on titan},    
pdfauthor={Sonal Kumar Jain},   
pdfsubject={Cameron band and UV doublet band},   
pdfcreator={TeXlive},   
pdfproducer={TeXmakerX}, 
pdfnewwindow=true,      
colorlinks=false,        
linkcolor=blue,         
linkcolor=blue,         
citecolor=blue,        
filecolor=magenta,      
urlcolor=red            
}

\newcommand{\nt}{{$\mathrm{N_2}$}}
\newcommand{\ch}{{$\mathrm{CH_4}$}}
\newcommand{\dgr}{{$^\circ$}}

\journal{Icarus}

\begin{document}

\begin{frontmatter}



\title{\bf Production of N$_2$ Vegard-Kaplan and  other triplet band
 emissions in the dayglow of Titan}
\author{Anil Bhardwaj\corref{cor1}} 
\cortext[cor1]{Corresponding author. Fax: +91 471 2706535 } 
\ead{anil\_Bhardwaj@vssc.gov.in; bhardwaj\_spl@yahoo.com}
\author{Sonal Kumar Jain}
\ead{sonaljain.spl@gmail.com}
\address{Space Physics Laboratory,
Vikram Sarabhai Space Centre,
Trivandrum~695022, India}

\begin{abstract}
Recently the Cassini Ultraviolet Imaging Spectrograph (UVIS) has revealed
the presence of N$_2$ Vegard-Kaplan (VK) band ($ A^3\Sigma_u^+ - X^1\Sigma^+_g $)
emissions in Titan's dayglow limb observation. We present model calculations
for the production of various N$_2$ triplet states (viz., $A^3\Sigma_u^+,\,  B^3\Pi_g,\,
C^3\Pi_u,\, E^3\Sigma_u,\,  W^3\Delta_u$, and $B'^3\Sigma_u$ ) in the upper atmosphere of Titan.
The Analytical Yield Spectra technique is used to calculate steady state
photoelectron fluxes in Titan's atmosphere, which are in agreement with those
observed by the Cassini's CAPS instrument. Considering direct
electron impact excitation, inter-state cascading, and quenching ef{}fects, the
population of dif{}ferent levels of \nt\ triplet states are calculated under
statistical equilibrium.  Densities of all vibrational levels of
each triplet state and volume production rates for
various triplet states are calculated in the model.
Vertically integrated overhead
intensities for the same date and lighting conditions as the reported by UVIS observations
for \nt\ Vegard-Kaplan ($ A^3\Sigma_u^+ - X^1\Sigma^+_g $), First Positive
($ B^3\Pi_g - A^3\Sigma^+_u $), Second Positive ($ C^3\Pi_u - B^3\Pi_g $),
Wu-Benesch ($W^3\Delta_u - B^3\Pi_g$), and Reverse First Positive bands of N$_2$
are found to be 132, 114, 19, 22, and 22 R, respectively. Overhead intensities 
are calculated for each vibrational transition of all the triplet band emissions of \nt, which
span a wider spectrum of wavelengths from ultraviolet to infrared.
The calculated limb intensities of total and prominent transitions of VK band are presented.
The model limb intensity of VK emission within the 150--190 nm 
wavelength region is in good agreement with the
Cassini UVIS observed limb profile. An assessment of the impact of solar EUV
flux on the N$_2$ triplet band emission intensity has been made by using three
dif{}ferent solar flux models, viz., Solar EUV Experiment (SEE), SOLAR2000 (S2K)
model of \cite{Tobiska04}, and HEUVAC model of \cite{Richards06}. The calculated N$_2$ 
VK band intensity at the peak of limb intensity due to S2K and HEUVAC solar 
flux models is a factor of 1.2 and 0.9, respectively, of that obtained using
SEE solar EUV flux. The ef{}fects of higher \nt\ density and solar zenith angle
on the emission intensity are also studied.
The model predicted N$_2$ triplet band intensities during moderate (F10.7 = 150)
and high  (F10.7 = 240) solar activity conditions, using SEE solar EUV flux,
are a factor of 2 and 2.8, respectively, higher than those during solar minimum (F10.7 = 68)
condition.
\end{abstract} 

\end{frontmatter}

\section{Introduction}
\label{sec:int}
The Saturnian satellite Titan, the second biggest satellite in the solar system, is 
in many ways the closest analogue to Earth. Like Earth, Titan's atmosphere is
dominated by \nt. Hence, it is natural to expect that Titan's airglow will be dominated 
by emissions of \nt\ and its dissociation product N. In addition to \nt, Titan  also
contains a few percent  \ch\ in its atmosphere, with a mixing ratio of about 3\% 
near 1000 km altitude \citep{DeLaHaye07, Strobel09}.

The Voyager 1 Ultraviolet Spectrometer (UVS) provided the first ultraviolet (UV) airglow 
observation of Titan in the 53--170  nm band \citep{Broadfoot81}. The extreme
ultraviolet
spectrum was dominated by emissions near 95--99 nm, which were attributed
to  \nt\ Carroll-Yoshino (CY) $c_4^{'1}\Sigma_u^+$ -- $X^1\Sigma_g^+$ (0, 0) and (0, 1)
Rydberg bands \citep{Strobel82}. Far ultraviolet emissions present
were LBH bands of \nt, and N and N$^+$ lines \citep{Broadfoot81,Strobel82}. By
employing multiple scattering model for CY band emissions, \cite{Stevens01} showed that CY
(0--0) should be weak and
undetectable, while CY (0--1) should be prominent emission at 981 nm and the features
at 950 nm are N I lines. Thus, there is no need to invoke magnetospheric electron
impact excitation \citep{Stevens01}.

After Voyager UVS, Cassini Ultraviolet Imaging Spectrograph (UVIS) provided
the next observation of Titan's airglow in the extreme ultraviolet (EUV, 56.1--118.2 nm)
and far ultraviolet (FUV, 115.5--191.3 nm) wavelengths \citep{Ajello07,Ajello08}.
These disk observations of Titan on 13 Dec. 2004 showed the presence of \nt\ LBH bands, atomic
multiplets of NI and N$^+$ lines, and  features
at 156.1 and 165.7 nm reportedly from CI \citep{Ajello08}.
Recently, limb observation of Titan by UVIS obtained on 22 June 2009 has revealed 
the presence of \nt\ Vegard-Kaplan (VK) ($ A^3\Sigma_u^+ - X^1\Sigma^+_g $) bands in
the FUV spectrum \citep{Stevens11}. Also, no CI emissions are reported to be observed.
\cite{Stevens11} showed that model emissions in the 150--190 nm VK band are
consistent with UVIS observations.

The \nt\ VK bands are a common feature in \nt\ atmospheres and have been studied
extensively on Earth \citep[e.g.,][]{Cartwright78,Meier91,Broadfoot97}. The \nt\ VK bands
have been observed recently on Mars by SPICAM aboard Mars Express
\citep{Leblanc06,Leblanc07,Jain11}. These emission can also be observed on
Venus by SPICAV onboard Venus Express \citep{Bhardwaj11b}, but the bright sunlit limb
causes a problem in resolving the dayglow emission from scattered light from clouds.

This paper presents a detailed model calculation for the production of \nt\ triplet
band emissions on Titan, including the recently observed \nt\ VK bands. The model
includes interstate cascading and quenching, and uses the Analytical Yield Spectra
approach for the  calculation of electron impact excitation of triplet bands, and
is similar to the model used for studying the \nt\ triplet emissions on Mars 
\citep{Jain11} and Venus \citep{Bhardwaj11b}. We also present the overhead emission
intensities of triplet bands, which lie in ultraviolet, visible, and
infrared  wavelengths. The calculated limb profile of \nt\ VK 150--190 nm
emission is compared with the Cassini UVIS observation. Impact on the intensity of 
\nt\ triplet emissions due to  changes in the solar EUV
flux model, solar activity, and Titan's \nt\ density are discussed.

The \nt\ triplet band emissions span a wide spectrum of electromagnetic radiation
covering EUV-FUV-MUV, visible, and infrared \citep{Jain11,Bhardwaj11b}.
Major emissions in \nt\ VK band lie in the wavelength range 200-400 nm, and a few significant
emissions in the visible. \nt\ triplet
First Positive ($ B \rightarrow A $), Wu-Benesch ($ W \rightarrow B $), and $ B' \rightarrow B $
bands have prominent emissions in the infrared region. Thus, beside observations
of Titan's dayglow by the Cassini UVIS in EUV and FUV region, the Cassini Visual and
Infrared Mapping Spectrometer (VIMS), which has a wide spectral range 300--5100 nm
\citep{Brown04} and Imagining Science Subsystem (ISS, 250--1100 nm) \citep{Porco04},
might be able to detect some of the bright emissions of \nt\
triplet bands in the MUV, visible, and infrared wavelengths predicted by our model. The model 
calculations presented in the paper would also be useful for any \nt-containing 
planetary atmospheres.

\section{Model Input}
\label{sec:mi}
The \nt\ density profile in our model is based on the observation made by Huygens
Atmospheric Structure Instrument (HASI) on-board Huygens probe \citep{Fulchignoni05}.
Following the approach of \cite{Stevens11}, the density of \nt\ is reduced by a factor
of 3.1 to bring the HASI \nt\ density at 950  km  to the level of measured density by Ion and
Neutral Mass Spectrometer (INMS)
\citep{DeLaHaye07} aboard the Cassini spacecraft. Since reduction in \nt\ density af{}fects
the altitude of peak production of \nt\ triplet bands, the  ef{}fect
of higher \nt\ density on emission intensities is discussed in  the 
Section~\ref{subsec:atm-ef{}fect}. Density of \ch\ in our model is based on the
UVIS stellar occultation experiment reported by \cite{Shemansky05}.

Photoabsorption and photoionization cross sections of \nt\ and \ch\ are taken from
photo-cross sections and rate coef{}ficients database (\url{http://amop.space.swri.edu})
\citep{Huebner92}. The branching ratios for excited states of N$_2^+$ and 
CH$_4^+$  are taken from \cite{Avakyan98}. Franck-Condon factors and transition 
probabilities required for calculating the intensity of a specific band $\nu'-\nu''$
of \nt\ are taken from \cite{Gilmore92}. Inelastic cross sections for the electron
impact on \nt\ are taken from \cite{Jackman77}, except for the triplet states,
which are taken from \cite{Itikawa06}, and fitted analytically for ease of usage in
the model \citep{Jackman77, Bhardwaj09,Bhardwaj11b,Jain11}. The fitted parameters
are given in \cite{Jain11}.

The solar EUV flux is a crucial input required in modelling the upper atmospheric
dayglow emissions. The solar EUV flux directly controls the photoelectron production rate,
and  hence the intensity of emission in the planetary atmosphere that is produced by 
electron impact excitation, like \nt\ triplet band emissions, since all transitions between the
triplet states of \nt\ and the ground state are spin forbidden. We have used the solar
irradiance measured at Earth (between 2.5 to 120.5 nm) by Solar EUV Experiment
(SEE, Version 10.2) \citep{Woods05,Lean11} on 23 June 2009 (F10.7 = 68) at 1 nm spectral
resolution. The solar flux has been scaled to the Sun-Titan distance (9.57 AU) to
account for the weaker flux on Titan. To evaluate the impact of solar EUV flux
model on emission intensities we have also used solar EUV flux from SOLAR2000 (S2K)
v.2.36 model of \cite{Tobiska04} and HEUVAC solar EUV flux model of \cite{Richards06}
for the same day. All calculations are made at solar zenith angle of 60\dgr\ unless
otherwise mentioned in the text.

\begin{figure}[h]
\noindent\includegraphics[width=20pc]{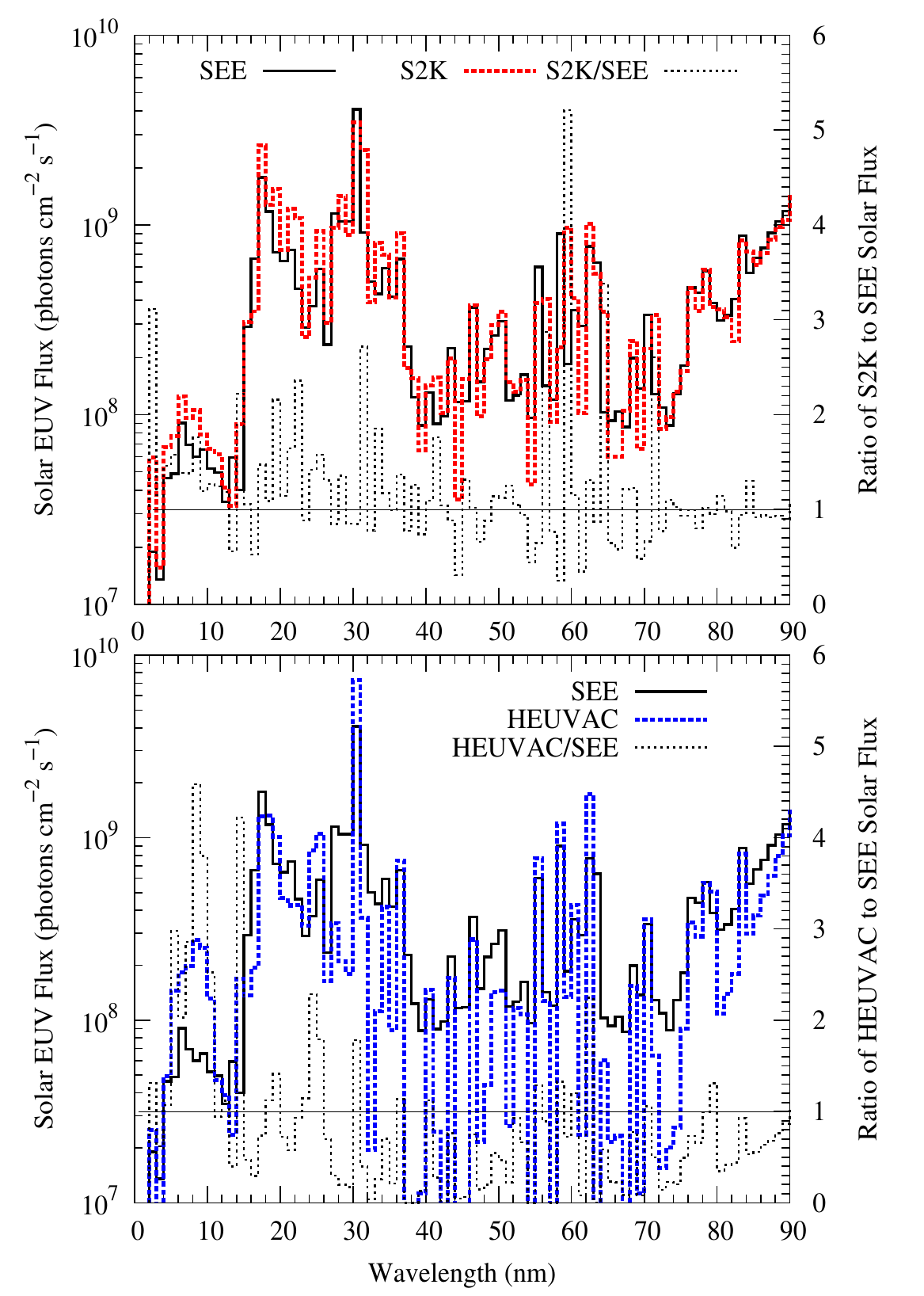} 
\caption{Comparison of SEE, S2K, and HEUVAC solar EUV flux models on 23 June
2009 at 1 AU. (top) SEE solar EUV flux compared with S2K. (bottom) SEE solar EUV flux comparison with 
HEUVAC. The ratio of solar EUV fluxes is also shown with magnitude on right side Y-axis.
Thin solid horizontal line depicts the S2K/SEE and HEUVAC/SEE solar flux ratio = 1.}
\label{fig:solar-flux}
\end{figure}
Figure~\ref{fig:solar-flux} (top panel) shows the solar EUV fluxes at 1 AU generated using
SEE and S2K model on 23 June 2009. Ratio of S2K and SEE solar EUV flux is also 
shown in the figure. Above 60 nm the SEE and S2K model solar EUV
fluxes are in general agreement with each other, but below 60 nm the solar flux
of S2K model is higher than that from SEE model. Higher EUV flux at shorter wavelengths
would result in larger number of higher energy photoelectrons; hence, volume production rates
calculated using S2K solar EUV flux is expected to be higher than that obtained using
SEE solar EUV flux (see Section~\ref{subsec:sf-ef{}fect}).

Figure~\ref{fig:solar-flux} (bottom panel) shows the comparison of SEE solar EUV flux
with the solar flux calculated using HEUVAC model, with both the daily F10.7 and the
F10.7--81 days average values set to 68 (in units of $10^{-22}\, W \, m^{-2} \, Hz^{-1}$),
conditions appropriate for the date of the UVIS limb observations reported by \cite{Stevens11}. Since the
HEUVAC model provides solar flux up to 105 nm only, the flux in the 105--120.5 nm range is
assumed the same as that in the SEE model. Since the solar flux at higher ($>$ 105 nm)
wavelengths does not contribute to the photoelectron production,
the inclusion of SEE solar flux in the HEUVAC model at wavelengths higher than 105 nm
would not af{}fect our calculation results. At wavelength above 30 nm HEUVAC solar
fluxes are smaller than those of SEE model, but at shorter ($<$ 30 nm) wavelengths HEUVAC
fluxes are higher than SEE model fluxes. At  a few shorter ($<$ 15 nm) wavelengths  HEUVAC
solar fluxes are as high as a factor of 4 compared to SEE solar flux. The photoelectrons
produced due to larger solar fluxes at shorter wavelengths in HEUVAC model would result
in larger number of higher energy photoelectrons that can go deeper in the atmosphere,
hence a significant lowering of the peak of volume production rate is
expected when the HEUVAC solar EUV flux is used (see following section).

\section{Results and discussion}
\label{sec:RD}
\subsection{Photoelectron Flux}
\label{subsec:pef}
To calculate the photoelectron flux we have adopted the Analytical Yield 
Spectra (AYS) technique 
\citep[cf.][]{Singhal84,Bhardwaj90a,Bhardwaj90b,Bhardwaj96,Singhal91,Bhardwaj99a,
Bhardwaj03,Bhardwaj99d,Bhardwaj99b}. The AYS is the analytical representation of
numerical yield spectra obtained using the Monte Carlo model 
\citep[cf.][]{Singhal80,Singhal91,Bhardwaj99d,Bhardwaj99b,Bhardwaj09}. Using AYS
the photoelectron flux has been calculated as \citep[e.g.,][]{Singhal84,Bhardwaj99b}

\begin{equation}\label{eq:a}
\phi(Z,E)=\int_{W_{kl}}^{100} \frac{Q(Z,E) U(E,E_0)}
{{\displaystyle\sum_{l}} n_l(Z)\sigma_{lT}(E)} \ dE_0
\end{equation}
where $\sigma_{lT}(E)$ is the total inelastic cross section for the $l$th gas,
at energy $E$, $n_l(Z)$ is its density at altitude $Z$, $ W_{kl} $ is the threshold of
$ k $th excited state of gas $ l $, and $U(E,E_0)$ is the
two-dimensional AYS, which embodies the non-spatial information of electron degradation
process. It represents the equilibrium number of electrons per unit energy at an
energy $E$ resulting from the local energy degradation of an incident
electron of energy $E_0$. For the \nt\ gas it is given as \citep{Singhal80}

\begin{figure}
\includegraphics[width=20pc]{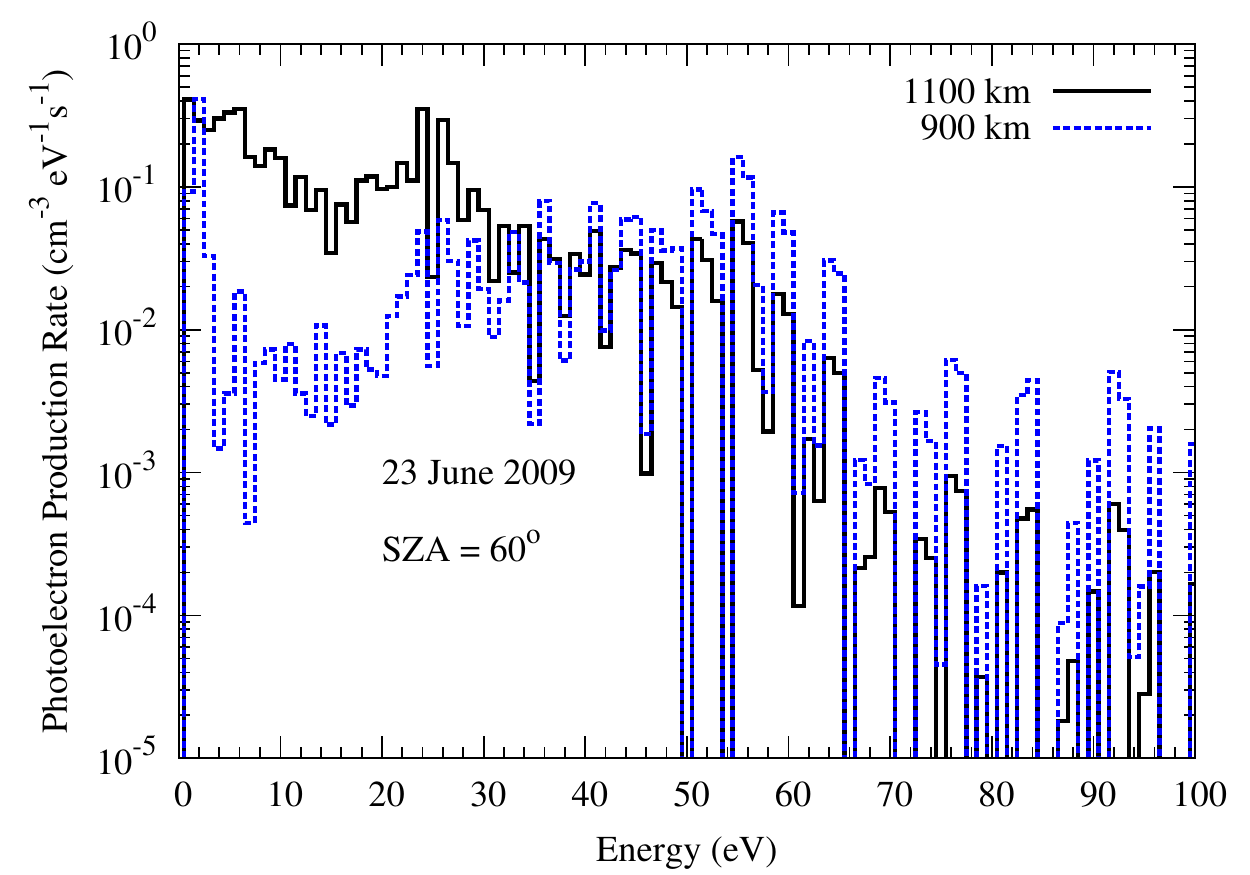}
\caption{Calculated photoelectron production spectrum at 1100 and 900 km at SZA = 60\dgr\ using SEE solar EUV
flux on 23 June 2009.}
\label{fig:pr}
\end{figure}
\begin{equation}\label{eq:d}
      U(E,E_0)=C_0+C_1(E_k+K)/[(E-M)^2+L^2].
\end{equation}
Here $C_0$, $C_1$, $K$, $M$, and $L$ are the
fitted parameters which are independent of the energy, and 
whose values are given by \cite{Singhal80}. The term $Q(Z,E)$ in 
equation~(\ref{eq:a}) is the primary photoelectron production rate 
\citep[cf.][]{Bhardwaj90a,Michael97,Bhardwaj03,Jain11}. Figure~\ref{fig:pr}
shows the calculated energy spectrum for photoelectron production at
900 and 1100 km on 23 June 2009 at SZA=60\dgr\ using SEE solar EUV
flux. Prominent peaks around 24--26 eV are due to the ionization of
\nt\ in dif{}ferent excited states by the solar He II Lyman-$ \alpha $ line at 303.78 \AA.
Photoelectron energy spectrum below 25 eV at altitude of 900 km is smaller than
that at 1100 km, since electrons below 25 eV mainly produced by solar EUV photons
$ > $ 30 nm which are attenuated at higher altitudes ($ > $ 900 km). Higher energy
photons can still penetrate deeper in the atmosphere and attain unit optical 
depth at lower altitudes ($ < $ 1100 km). That is why photoelectron production spectrum at
higher energies ($ > $ 50 eV) is higher at 900 km compared to electron spectrum 
at 1100 km.

During the degradation of photoelectrons in the atmosphere of Titan we
have not considered \ch\ since \nt\ is the dominant species. Contribution of
\ch, which is having a mixing ratio of $ \sim $3\% near 1000 km, to the
photoelectron flux is less than 10\% \citep{Stevens11}. The ef{}fect of omitting the
\ch\ contribution in the photoelectron production rate is less than 5\% in our
calculation. \cite{Stevens11} also have stated that neglecting the \ch\ contribution
in the calculation of photoelectron production rate results in only less than 5\%
enhancement in the EUV and FUV volume production rates for the UVIS observation
conditions.

\begin{figure}[h]
\includegraphics[width=20pc]{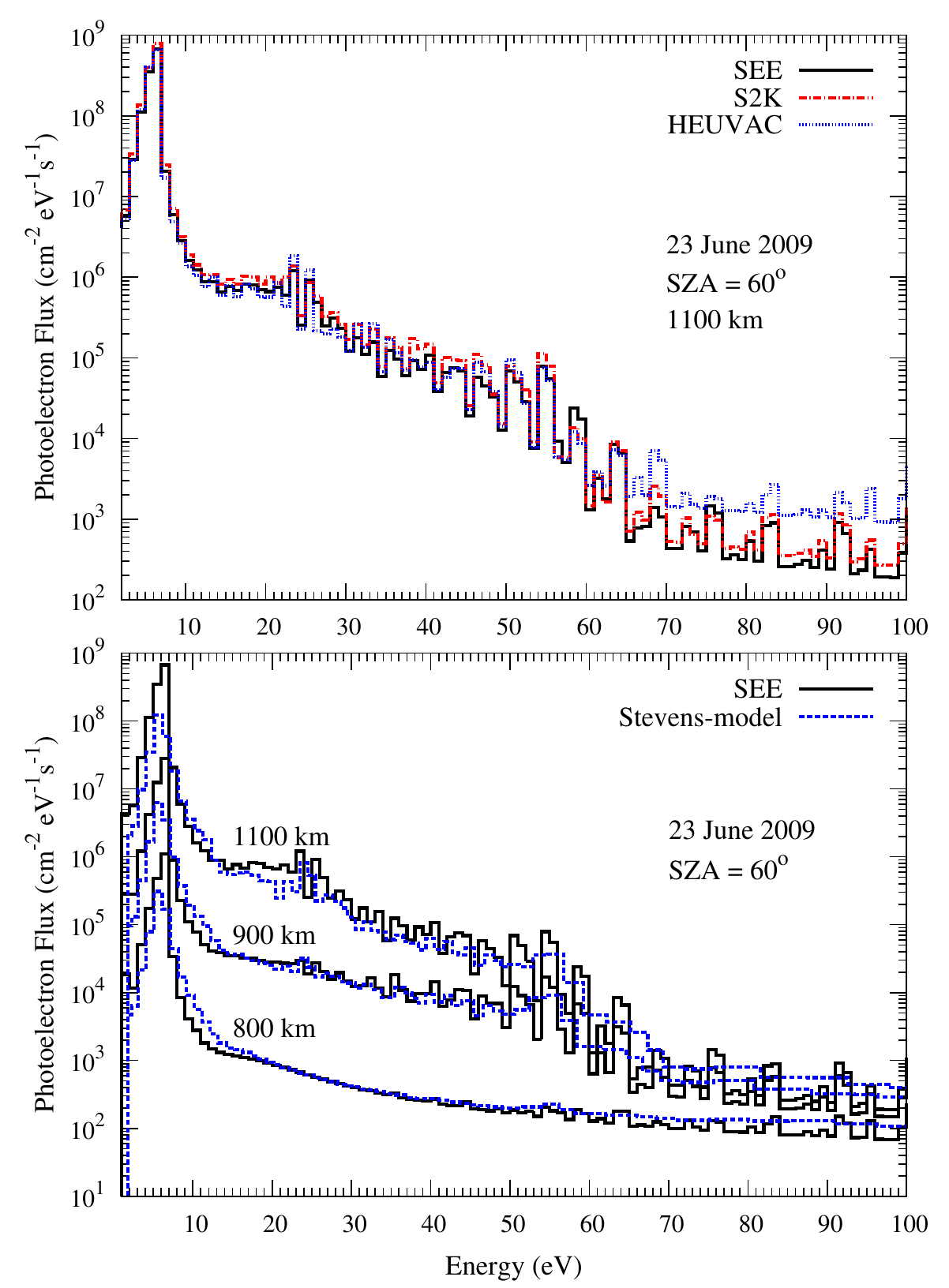}
\caption{(top panel) Model calculated photoelectron flux at 1100 km using 
SEE,  SOLAR2000 (S2K), and HEUVAC solar flux models. (bottom panel) Comparison of model calculated
photoelectron fluxes using SEE solar EUV flux at 800, 900, and 1100 km with those of \cite{Stevens11}.}
\label{fig:pef}
\end{figure}
Figure~\ref{fig:pef} (top panel) shows the steady state photoelectron flux at altitude of
1100 km and solar zenith angle 60\dgr. Around 6 eV, photoelectron flux is maximum
with a value of few $ 10^8$ cm$^{-2}$ s$^{-1}$ eV$^{-1}$.
Due to the degradation of electrons in the atmosphere, the photoelectron flux
is smoother compared to the photoelectron production energy spectrum (cf. Fig.~\ref{fig:pr});
still prominent peak can be seen in the flux, {\it e.g.}, peak at 24 eV is present in the
photoelectron flux. Photoelectron fluxes calculated by using S2K
and HEUVAC solar EUV fluxes are also shown in Figure~\ref{fig:pef}. 
As discussed in Section~\ref{sec:mi}, due to higher solar EUV flux in
the S2K model, the photoionization yield is slightly higher compared to that
in the SEE model, which is responsible for the higher photoelectron flux when
S2K model is used. Whereas at energies below 60 eV, photoelectron flux calculated
using HEUVAC model is lower than that calculated using SEE solar EUV flux, but
at higher energies ($>$60 eV) photoelectron flux calculated using HEUVAC model is
higher than that calculated using both SEE and S2K models. Higher
photoelectron flux at higher energies ($>$60 eV ) for HEUVAC model
is due to the higher solar EUV flux at shorter wavelengths
(cf. Figure~\ref{fig:solar-flux}).

Figure~\ref{fig:pef} (bottom panel) shows the steady state photoelectron fluxes 
at three dif{}ferent altitudes along with the calculated photoelectron fluxes of
\cite{Stevens11}. At and below 6 eV, our calculated photoelectron flux is higher
than that of \cite{Stevens11}. This is due to the lack of electron-electron collision loss
consideration in our model, which is an important 
electron energy loss process below 10 eV \citep{Bhardwaj90b,Bhardwaj11d}.
Between 6 and 15 eV, calculated photoelectron fluxes of \cite{Stevens11}
are higher than our calculated values at all altitudes. At 1100 km, our calculated
photoelectron flux between 15 and 25 eV is higher than that of
\cite{Stevens11}. This dif{}ference in the photoelectron flux between two model
calculations may be due to the slightly dif{}ferent treatment of the altitude dependence 
of electron degradation in both models. In our model, the AYS approach used for the
calculation of photoelectron flux is based on the Monte Carlo model
\citep[cf.][]{Singhal80,Singhal91,Bhardwaj93,Bhardwaj99d,Bhardwaj99b,Bhardwaj09},
while in the AURIC model \citep{Strickland99} it is based on the solution of Boltzmann equation.
Above 25 eV, the photoelectron flux calculated by
\cite{Stevens11} is consistent with our model values. The peak structures in our calculated
photoelectron flux are slightly dif{}ferent than the photoelectron flux of \cite{Stevens11}.
This dif{}ference is due to the dif{}ferent branching ratios used in both models. In our model,
stable branching ratios taken from \cite{Avakyan98} are used, whereas in AURIC model branching
ratios are wavelength dependent.

\begin{figure}
\includegraphics[width=20pc]{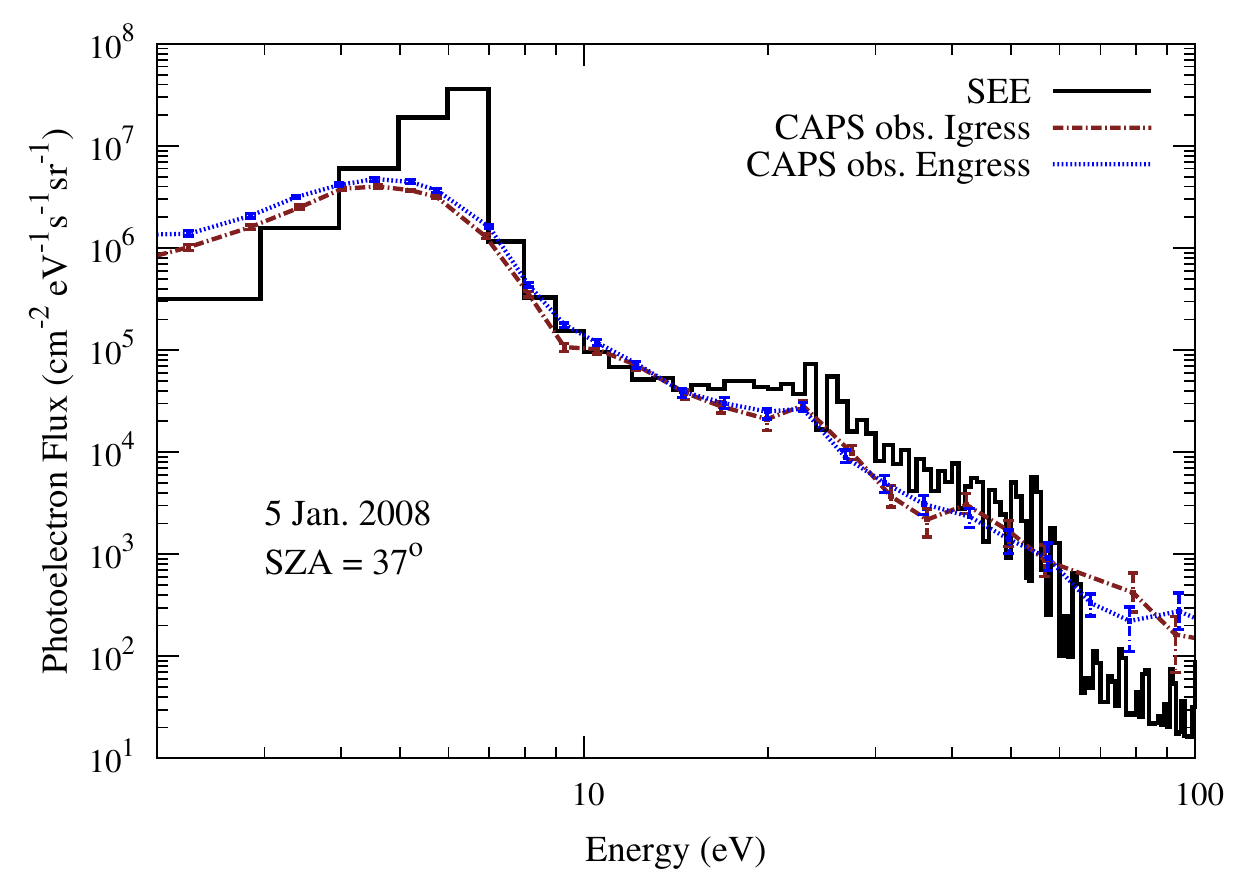}
\caption{Model calculated photoelectron flux on 5 Jan. 2008 at altitude of
1100 km obtained by using SEE solar EUV flux compared with the Cassini CAPS
observation taken from \cite{Lavvas11}.}
\label{fig:pef-cmp}
\end{figure}

To compare the calculated photoelectron flux with Cassini observations we have run
our model taking the HASI \nt\ density and SEE solar EUV flux 
on 5 January 2008 (F10.7 = 79.7) at SZA=37\dgr. Figure~\ref{fig:pef-cmp} shows the model
calculated photoelectron flux at 1100 km along with the photoelectron flux
observed by the CAPS instrument (energy resolution $ \Delta E/E = 16.7\% $) on-board
Cassini taken from \cite{Lavvas11}.
Model calculated photoelectron flux agrees well with the observed flux
between 7 and $ \sim $20 eV. Above 20 eV model predicted photoelectron flux
is slightly higher than the observation.
At higher energies ($ > $60 eV) the calculated photoelectron flux starts decreasing
sharply compared to the observed flux. \citeauthor{Lavvas11} have also observed similar
dif{}ferences in their calculated and the observed photoelectron flux at energies $ > $ 60 eV,
which they attributed to the instrument artifact \citep{Lavvas11, Arridge09}.

\subsection{Volume Emission rates}
\begin{figure}
\includegraphics[width=20pc]{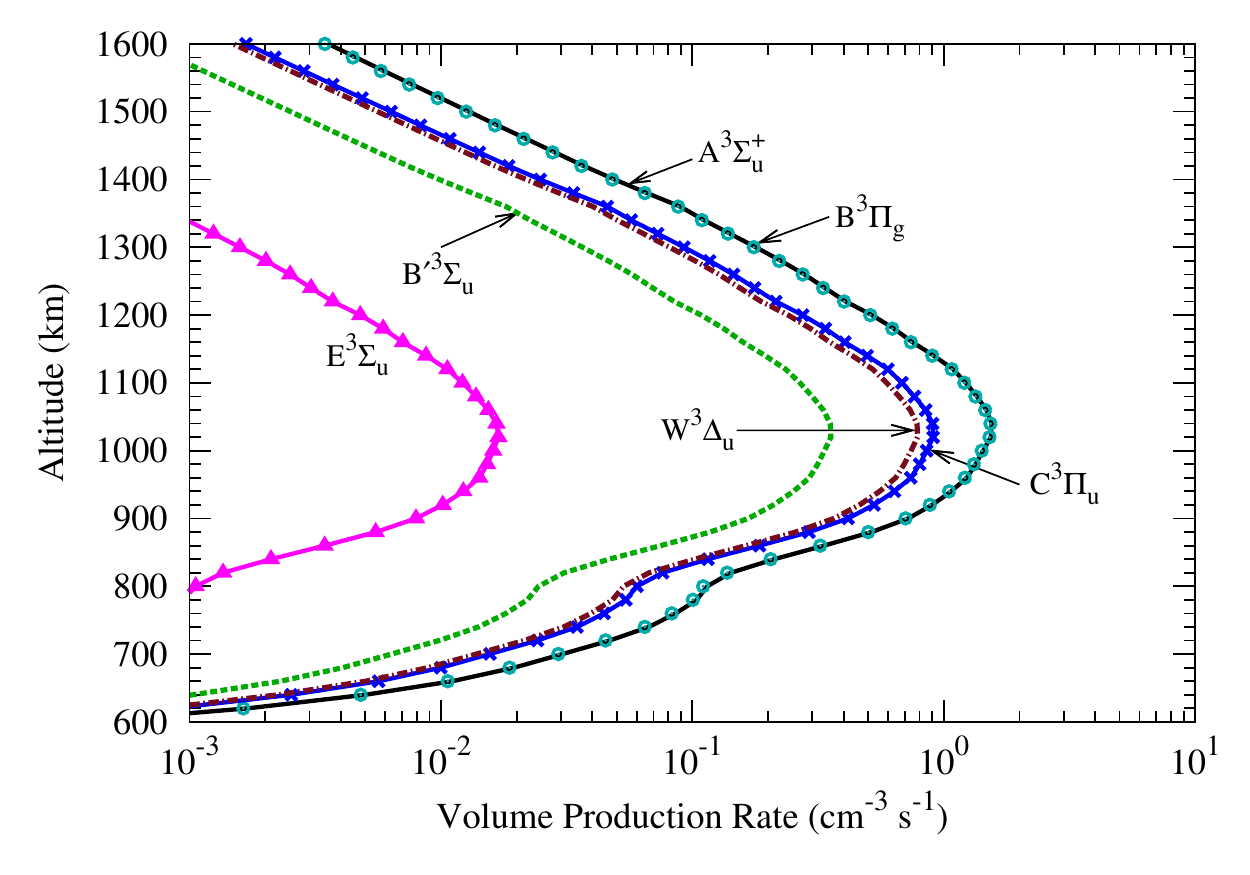}
\caption{Calculated volume production rate of dif{}ferent triplet states of \nt\ due to 
photoelectron impact at SZA=60\dgr\ for SEE solar EUV flux.}
\label{fig:ver-tri}
\end{figure}

Using the photoelectron flux $\phi(Z,E)$ obtained in equation~(\ref{eq:a}),
the volume excitation rate for \nt\ emissions is calculated as

\begin{equation}\label{eq:e}
V_i(Z) = n(Z) \int _{E_{th}}^{E} \phi(Z, E) \sigma_i(E) dE
\end{equation}
where $n(Z)$ is the density of \nt\ at altitude $ Z $ and  $ \sigma_i(E)$
is the electron impact cross section for the $i$th state at energy $E$, for which the
threshold is $E_{th}$. Figure~\ref{fig:ver-tri} shows the volume excitation rates
for \nt\ triplet states ($ A,\, B,\, C, \, W, \,B',$ and $ E $) by photoelectron
impact excitation. The production rates for all the states peak around 1025
km, which is $\sim$ 25 km higher than the calculated peak altitude of $ \sim $1000 km
of \cite{Stevens11}. This dif{}ference  in height of peak production might due to the
dif{}ferent treatment of altitude dependence of photoelectrons in the two models.

The \nt\ triplet $E$, $C$, $W$, and  $ B'$ states populate the $B$ state, which 
in turn radiate to the state $A$ (First Positive band). Further the interstate
cascading $ B^3\Pi_g \rightleftharpoons A^3\Sigma^+_u $ and $ B^3\Pi_g 
\rightleftharpoons W^3\Delta_u $ are also important in populating the $ B $ 
level \citep{Cartwright71, Cartwright78,Jain11,Bhardwaj11b}. To calculate
the total population of dif{}ferent vibrational levels of state $A$, we 
solve the equations for statistical equilibrium based on the formulation
of \cite{Cartwright78}. Contributions of cascading from higher triplet
states and interstate cascading and quenching by atmosphere constituents
are included in the calculation.
At a specified altitude, for a 
vibrational level $ \nu $  of a state $ \alpha $, the population is 
determined using the statistical equilibrium equation
\begin{equation}\label{eq:sta}
V^\alpha q_{0\nu}  + \sum\limits_{\beta}\sum\limits_{s}A^{\beta\alpha}_{s\nu}\, n^\beta_s
= \{ K^\alpha_{q\nu} + \sum\limits_{\gamma}\sum\limits_{r}A^{\alpha\gamma}_{\nu r} \}n^\alpha_{\nu}
\end{equation}
where, 
$ V^\alpha $ is electron impact volume excitation rate (cm$^{-3}$\ s$^{-1}$)
of state $\alpha$; $q_{0\nu}$ is Franck-Condon factor for the  excitation from ground
level to $\nu$ level of state $ \alpha $; $A^{\beta\alpha}_{s\nu}$ is
transition probability (s$^{-1}$) from state $ \beta(s)$ to $ \alpha(\nu)$;
$K^\alpha_{q\nu}$ is total electronic quenching frequency (s$^{-1}$) of 
level $\nu$ of state $ \alpha $ by  all the gases 
defined as: $\sum\limits_{l} K_{q(l)\nu}^\alpha \times n_{l}  $,
where, $K_{q(l)\nu}^\alpha$ is the quenching rate  coef{}ficient of level $ \nu $
of state $ \alpha $ by the gas $ l $ of density $ n_l $;
$ A^{\alpha\gamma}_{\nu r} $ is transition from level $ \nu $ of state $ \alpha $ 
to vibrational level $r$ of state $ \gamma $; $ n $ is density (cm$^{-3}$);
$ \alpha, \beta$, and   $\gamma $ are electronic states; and
$ s$ and  $r $ are  source and sink vibrational levels, respectively.

While calculating the cascading from  $ C $ state, we have accounted for
predissociation. The $ C $ state predissociates approximately half the time
(this is an average value for all  vibrational levels of the $ C $ state;
excluding $ \nu = $ 0, 1, which do not predissociate at all)
\citep[cf.][]{Daniell86}. In Earth's airglow the \nt(A) levels get
ef{}fectively quenched by atomic oxygen and the abundance of O increases with
increase in altitude. Titan atmosphere is \nt\ dominated with small amount of \ch.
The quenching rates for dif{}ferent vibrational levels of \nt\ triplet
states by \nt\ are adopted from \cite{Morrill96} and by \ch\ is
taken from \cite{Clark80}.

\begin{figure}
\includegraphics[width=20pc]{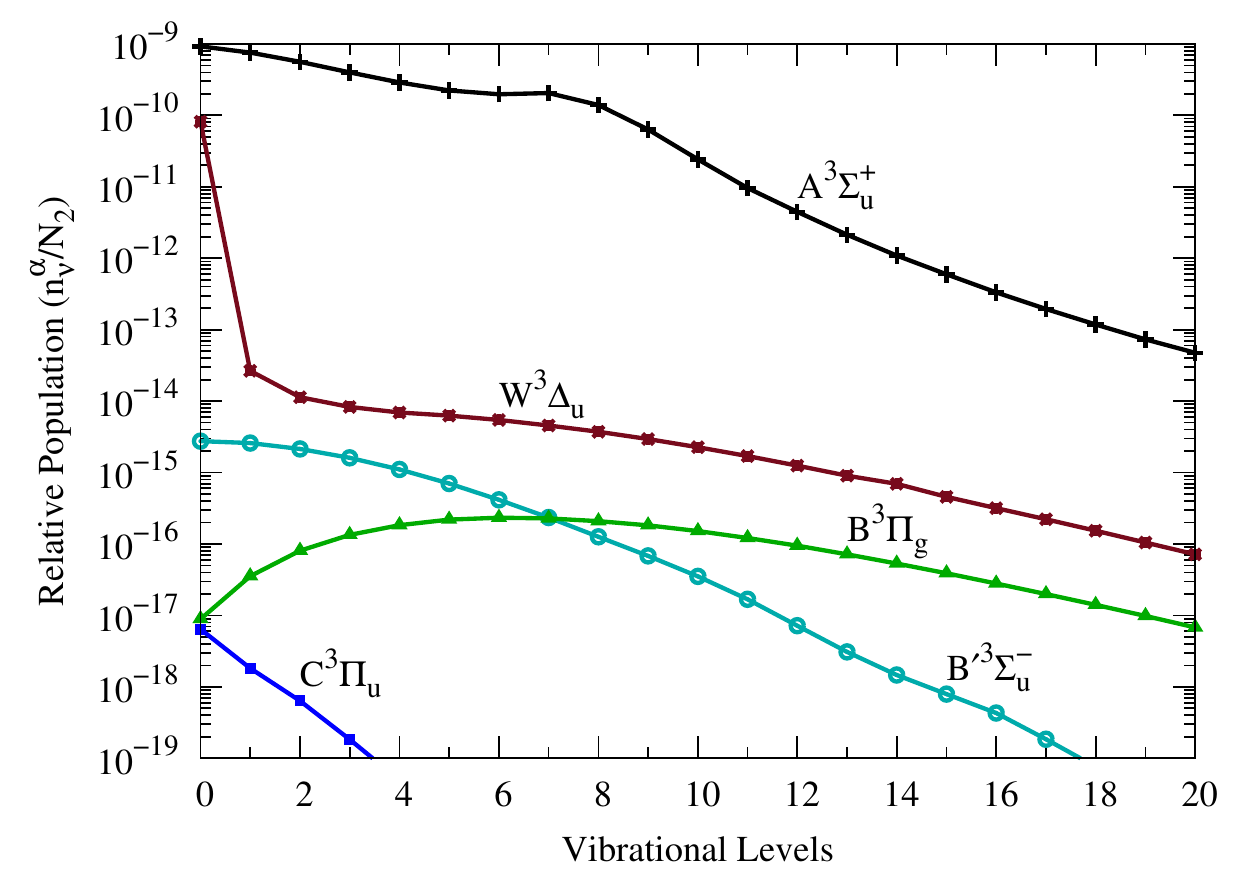}
\caption{Calculated relative vibrational population of the various triplet states of \nt\ at 
an altitude of 1100 km at SZA=60\dgr\ using SEE solar flux.}
\label{fig:vibpop}
\end{figure}

Figure~\ref{fig:vibpop} shows the population of dif{}ferent 
vibrational levels of triplet states of \nt\ relative 
to the ground state at 1100 km. We found that the ef{}fect of quenching 
is negligible on the vibrational population. \cite{Stevens11}
have taken the vibrational population of VK band up to 10
vibrational levels and the population of  vibrational
level 11 is taken as 44\% of that for $ \nu'=10 $.
In our model, the vibrational population is considered up to 
$ \nu'=20 $ levels. Our calculated population for $ \nu'=11 $ level
is $ \sim $40\% of the $ \nu'=10 $ level.

After calculating the steady state density of 
dif{}ferent vibrational levels of excited triplet states of 
\nt, the volume emission rate $ V_{\nu'\nu''}^{\alpha\beta} $ 
of a vibration band $\nu' \rightarrow \nu''$ can be obtained using
\begin{equation}\label{eq:g}
V_{\nu'\nu''}^{\alpha\beta} = n_{\nu'}^\alpha \times 
                            A_{\nu'\nu''}^{\alpha\beta} \quad (cm^{-3}\ s^{-1})
\end{equation}
where  $n_{\nu'}^\alpha$ is the density of vibrational 
level $\nu'$ of state $ \alpha $, and $A_{\nu'\nu''}^{\alpha\beta}$
is the transition probability (s$^{-1}$) for the transition 
from the $\nu'$ level of state $ \alpha $ to the $\nu''$ level of state $ \beta $.

\begin{figure}
\includegraphics[width=20pc]{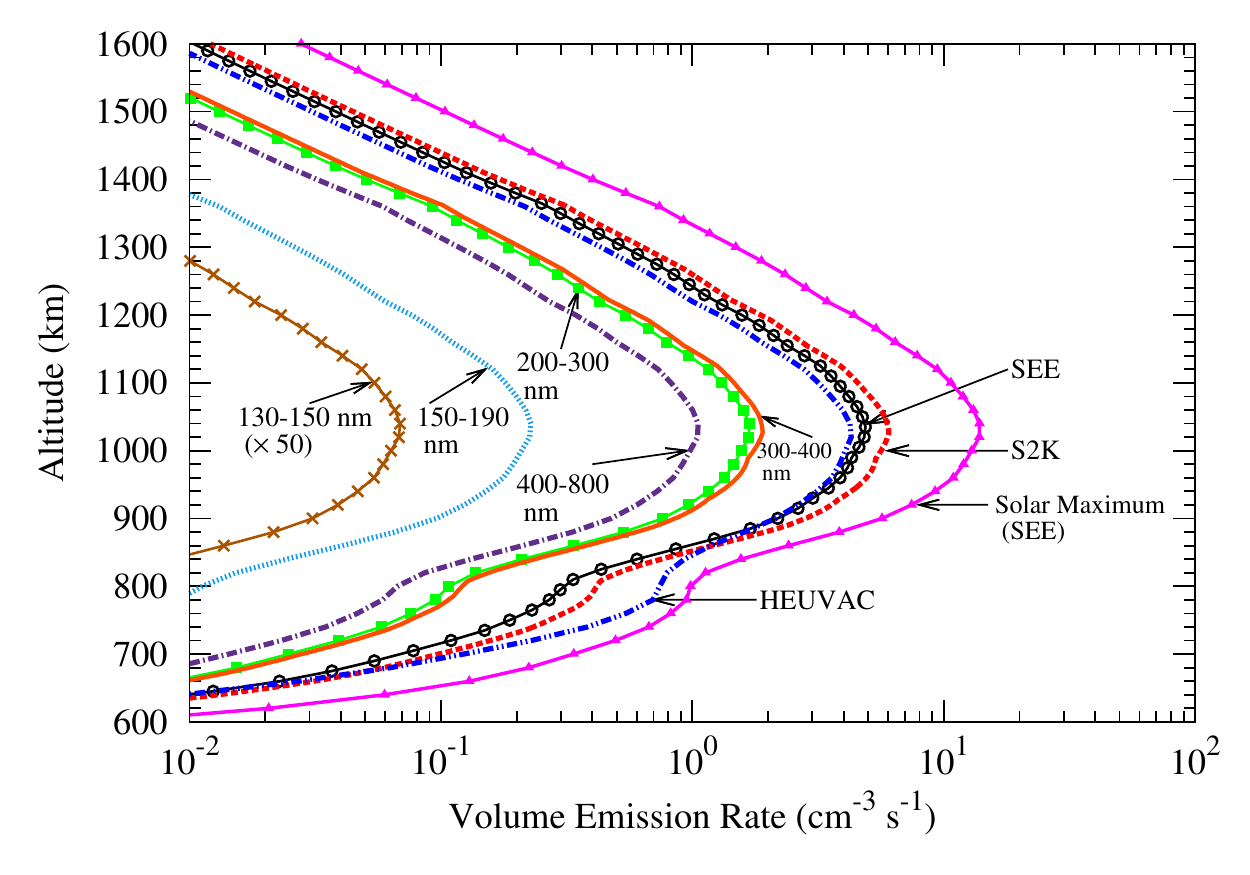}
\caption{Volume emission rate of total \nt\ VK band along with the emission rates of
VK band in  dif{}ferent wavelength regions calculated using
SEE solar flux model at SZA = 60\dgr. Total \nt\ VK band emission rate of 130--150 nm band 
is plotted after multiplying by a factor of 50. Emission rate 
profiles for SOLAR2000 (S2K) and HEUVAC model solar fluxes and in
solar maximum (for SEE solar EUV flux) are also shown.}
\label{fig:ver-cmp}
\end{figure}

Figure~\ref{fig:ver-cmp} shows the volume emission rates for the VK band.
The \nt\ VK band span wavelength range
from FUV to visible, and some transitions even emit at wavelength more
than 1000 nm. Figure~\ref{fig:ver-cmp} also shows the emission rates of 
VK bands in  FUV and visible wavelengths. Volume emission rates for VK bands
in the wavelength range 400--800, 300--400,  200--300,
and 150--200 nm are 22\%, 38\%, 35\%, and 4.5\% of the total VK band
emission rate. Contribution of VK band emissions in the  130--150 nm
wavelength range is very small (0.02\%).
In the visible and near infrared range (400--800 nm), the main
contribution comes from the emissions
between 400 and 500 nm, which comprises around 73\% of the 
VK visible emission band. Our calculated total VK band volume
emission rate is in good agreement with that of \cite{Stevens11}.

\begin{figure}
\includegraphics[width=20pc]{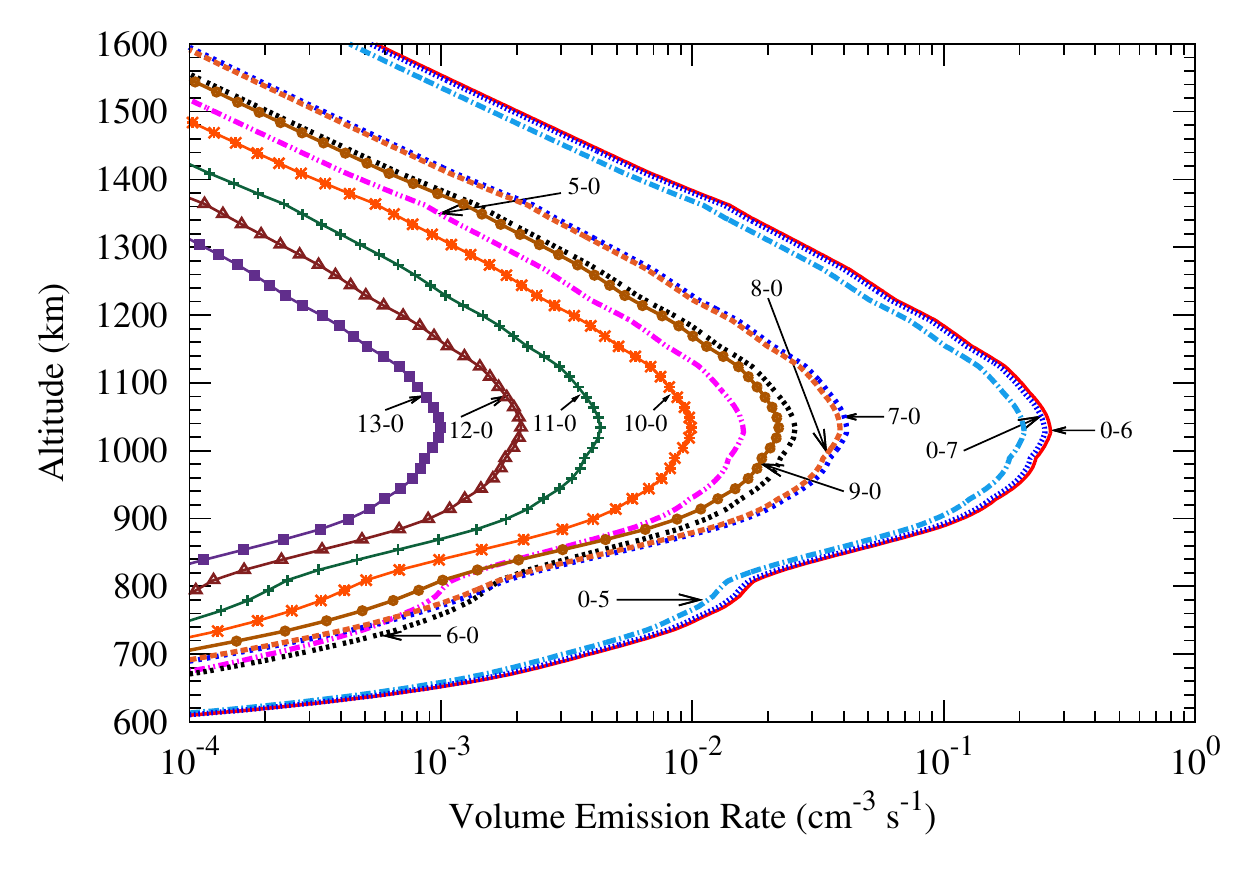}
\caption{Volume emission rate profiles of prominent transitions of \nt\ VK band emission
calculated using SEE solar flux model at SZA = 60\dgr.}
\label{fig:ver}
\end{figure}

Figure~\ref{fig:ver} shows the prominent VK band transition in 
ultraviolet region. VK (0, 5), (0, 6), and (0, 7) bands (between
200--300 nm) have been observed on Mars by SPICAM/Mars Express
\citep{Leblanc06,Leblanc07}. In the 150--200 nm region, VK (5, 0), (6, 0),
(7, 0), (8, 0), and (9, 0) bands are reported for the first time in the dayglow
of Titan \citep{Stevens11}. The production rates in these two 
wavelength band regions, 150--200 and 200--300 nm, dif{}fer by about an
order of magnitude but the altitude of peak production (1025 km) remains
the same for all the VK band emissions (Figure~\ref{fig:ver-cmp}).

The volume emission rates are vertically-integrated to calculate the overhead
intensities. Table~\ref{tab:oi-triplet} shows the total overhead intensity for
Vegard-Kaplan ($A \rightarrow X$), First positive ($B \rightarrow A$),
Second Positive ($C \rightarrow B$),  Herman--Kaplan ($E \rightarrow A$),
$E \rightarrow B$, Reverse First Positive ($A \rightarrow B$), and
$E \rightarrow C$ triplet bands of \nt\ at SZA=60\dgr. Since
the VK band spans a wide range of electromagnetic
spectrum, from FUV to visible wavelengths, we  also present in 
Table~\ref{tab:oi-triplet} the overhead intensities in dif{}ferent wavelength
regions of VK bands. Emissions in the 300--400 nm constitute a major fraction
of the total VK band emission followed closely by emissions in the
200--300 nm band, with contributions of around 38\% and
35\%, respectively. The 150--200 nm emission band contributes around 4.5\% to 
the total VK band intensity. Contribution of visible wavelength region
(400--800 nm) is also significant (22\%) in the total VK band intensity,
in which wavelength region 400--500 nm contributes $ \sim $16\% of the
total VK band intensity or 73\% of total visible band emission.

Table~\ref{tab:vk-oi} shows the overhead intensity  for all the
vibrational levels of \nt\ VK bands calculated using SEE solar EUV flux on 
23 June 2009 at SZA = 60\dgr. The VK (0, 6) emission (at 276.2 nm)
is the strongest emission in the VK band system having an overhead
intensity of $ \sim $7 R, which is around 5\% of the total VK band 
intensity and comprises around 15\% of VK band emissions in the 200--300 nm range.
The VK (0, 6) band has been observed on Mars \citep{Leblanc07,Jain11}.
In the dayglow spectrum of Titan VK (7, 0) transition is the strongest
emission observed by Cassini UVIS. For the VK (7, 0) band the model calculated
overhead intensity is 1.1 R, which is  0.8\% of the total VK band intensity.

The calculated overhead intensities of \nt\ First Positive (1P) transitions
are presented in Table~\ref{tab:BA-oi}. Prominent transitions in this
band lies above 600 nm. The 1P (1, 0) emission  at 888.3 nm is
the brightest followed by (0, 0) emission at 1046.9 nm, which
contribute around 13\% and 9\%, respectively, to the total
1P emission. Emissions between 600 and 800 nm wavelength 
consist of about 50\% of the total 1P band system. The calculated
overhead intensities of Second Positive (2P) band transitions are
presented in Table~\ref{tab:CB-oi}. Major portion of 2P band
emission lies in wavelengths between 300 and 400 nm, which
is more than 90\% of the total 2P band overhead intensity. Prominent
emissions in the 2P band system are (0, 0), (0, 1), (0, 2), and
(1, 0) transitions, having  overhead intensities of around
6.5, 4.5, 1.8, and 1.7 R, thus contributing around 34, 24, 9,
and 9\%, respectively, to the total 2P emission.

Tables~\ref{tab:WB-oi} and \ref{tab:BB-oi} show the calculated 
overhead intensities of Wu-Benesch ($ W \rightarrow B $) and 
$ B' \rightarrow B $ band emissions, respectively. Most of
the emissions in $ W \rightarrow B $ band are in infrared 
region with a little or negligible contribution 
from emissions below 800 nm. Similar is the case in 
$ B' \rightarrow B $ band system. Table~\ref{tab:EABC-oi} shows
the calculated overhead intensities of 
Herman--Kaplan ($E \rightarrow A$), $E \rightarrow B$,
and $E \rightarrow C$ bands of \nt, and Table~\ref{tab:r1p-oi}
shows the overhead intensities of Reverse First Positive (R1P) band
emissions. Prominent emissions in the R1P band system are in
infrared region, with (9, 0) emission being the strongest
having the overhead intensity of 1.9 R, which is around 
9\% of the total R1P emission.

\cite{Stevens11} suggested that \nt\ VK (8, 0) emission near 165.4 nm and
(11, 0) band near 156.3 nm could have been misidentified as
CI 165.7  and 156.1 nm emissions  by \cite{Ajello08}. Also, the VK (10, 0)
band could be the emission near 159.2 nm, which is reported as mystery
feature by \cite{Ajello08}.
We calculated the overhead intensities of  CI 165.7  and 156.1 nm emissions
due to electron impact dissociative excitation of \ch\ by using the emission
cross sections from \cite{Shirai02}. The model calculated overhead intensities
of  CI 156.1 and 165.7 nm emissions are $ 1.6 \times 10^{-3} $ and  $ 3.7
\times 10^{-3} $ R, respectively, which are about 2 orders of magnitude lower
than the VK (8, 0) and VK (11, 0) bands intensities (see Table~\ref{tab:vk-oi}).
We also estimated the solar scattered intensities for CI 165.7 and 156.1 nm
using the density of atomic carbon from the model of \cite{Krasnopolsky10}
and {\it g}-factor values of $7.21 \times 10^{-6}$ and $2 \times 10^{-5}$
sec$ ^{-1} $, respectively, at 1 AU. The overhead intensities of CI 156.1
 and 165.7 nm  due to solar fluorescence are an order of magnitude lower 
($1.98 \times 10^{-4}$ and $5.5\times 10^{-4}$ R, respectively) than that due to
photoelectron excitation. The dif{}ference of 2 to 3 orders of magnitude between
intensities of (8, 0) and (11, 0) VK bands and CI line emissions suggest that
bands near 156.1 and 165.4 nm might have been misidentified by \cite{Ajello08},
as reported by \cite{Stevens11}.

\begin{figure}
\includegraphics[width=20pc]{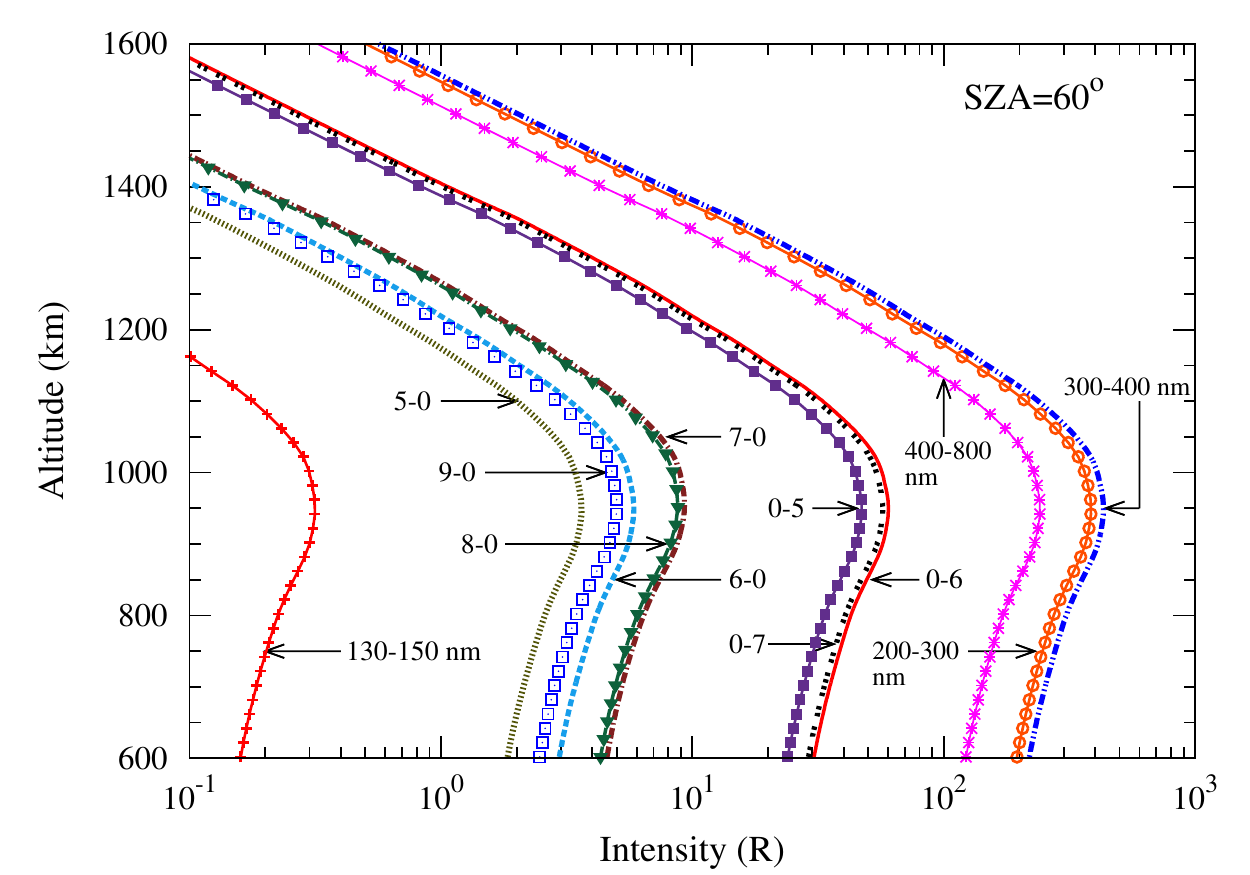}
\caption{Limb profiles of prominent transitions of \nt\ VK bands calculated
using SEE solar flux model at SZA = 60\dgr. Limb profiles of \nt\
VK band in dif{}ferent wavelength regions (130--150, 200--300, 300--400, and 400-800 nm)
are also shown.}
\label{fig:int-bands}
\end{figure}

\begin{figure}
\includegraphics[width=20pc]{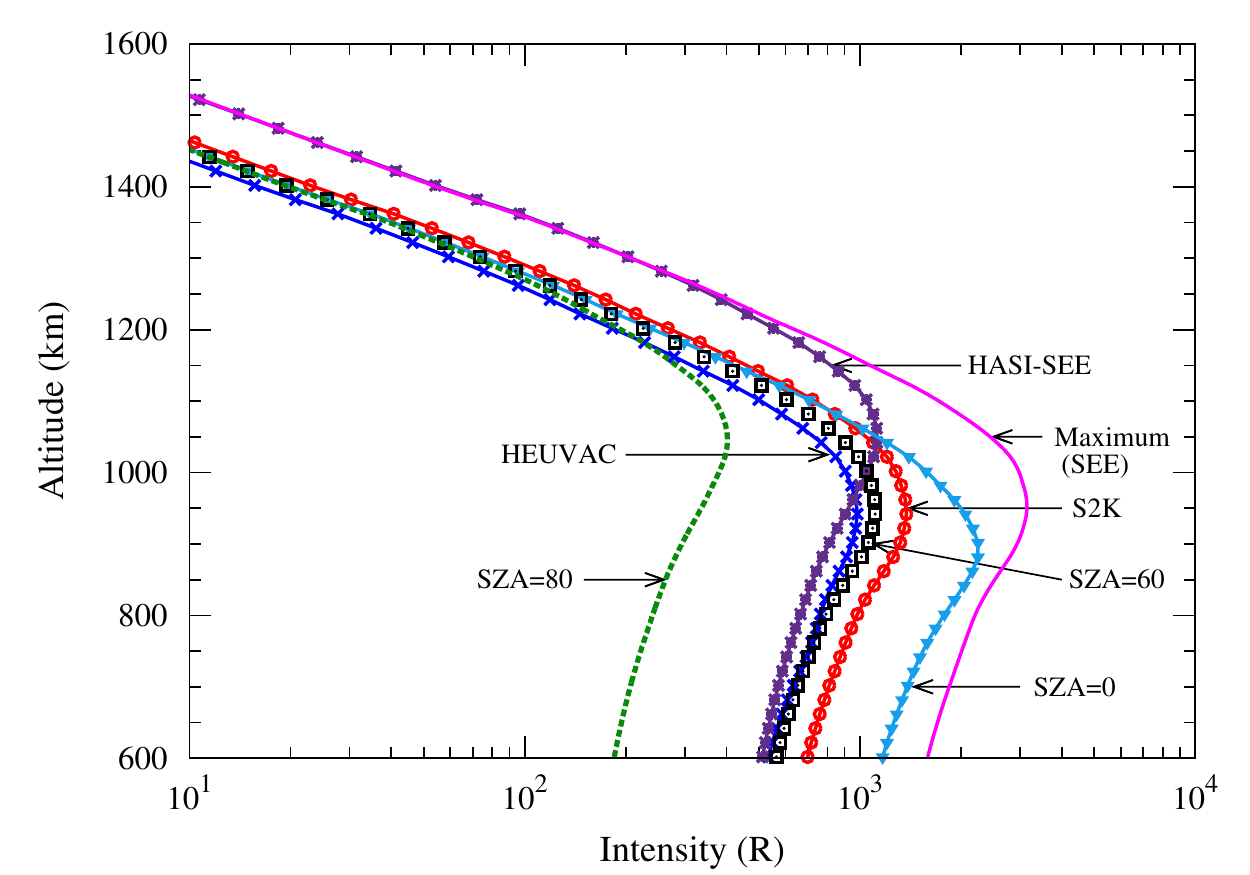}
\caption{Limb profiles of total \nt\ VK bands at three dif{}ferent solar zenith angles
(0, 60, and 80 degrees) calculated
using SEE solar EUV flux model. Limb profiles calculated using S2K and HEUVAC models,
and in solar maximum condition (for SEE solar flux) are shown for SZA = 60\dgr. 
Limb intensity profile calculated using HASI \nt\ density and SEE solar EUV flux
is also shown for SZA = 60\dgr.}
\label{fig:int-vk}
\end{figure}

The calculated band emission rate is integrated along the line of 
sight at a projected distance from the centre of Titan to obtain limb 
profile. Figure~\ref{fig:int-bands} shows the calculated limb intensities
of the prominent VK band emissions in ultraviolet region. Limb intensities
of VK bands in dif{}ferent wavelength regions are also shown in 
Figure~\ref{fig:int-bands}. The altitude of maximum limb intensity is
around 950 km for all the transitions, slightly higher than the calculated
\nt\ VK peak of \cite{Stevens11} of $ \sim $928 km. Figure~\ref{fig:int-vk} shows the limb
intensity of total VK band, which peaks at $ \sim $950 km, with a value of
around 1.1 kR. Limb intensities calculated at dif{}ferent solar zenith angle
are also shown in Figure~\ref{fig:int-vk}. The main ef{}fect of SZA is on the
altitude of peak limb intensity and intensity at the peak;
lower the value of SZA, the deeper the peak of the limb profile  with higher
intensity, which is due to the penetration of solar EUV at lower
altitudes in the atmosphere. For SZA = 0\dgr, the calculated peak limb intensity is
2.2 kR at an altitude of 892 km, whereas at SZA = 80\dgr, the peak limb intensity is
0.4 kR at an altitude of 1050 km.  Above 1200 km the ef{}fect of solar
zenith angle is not seen in the limb intensities.

\begin{figure}
\includegraphics[width=20pc]{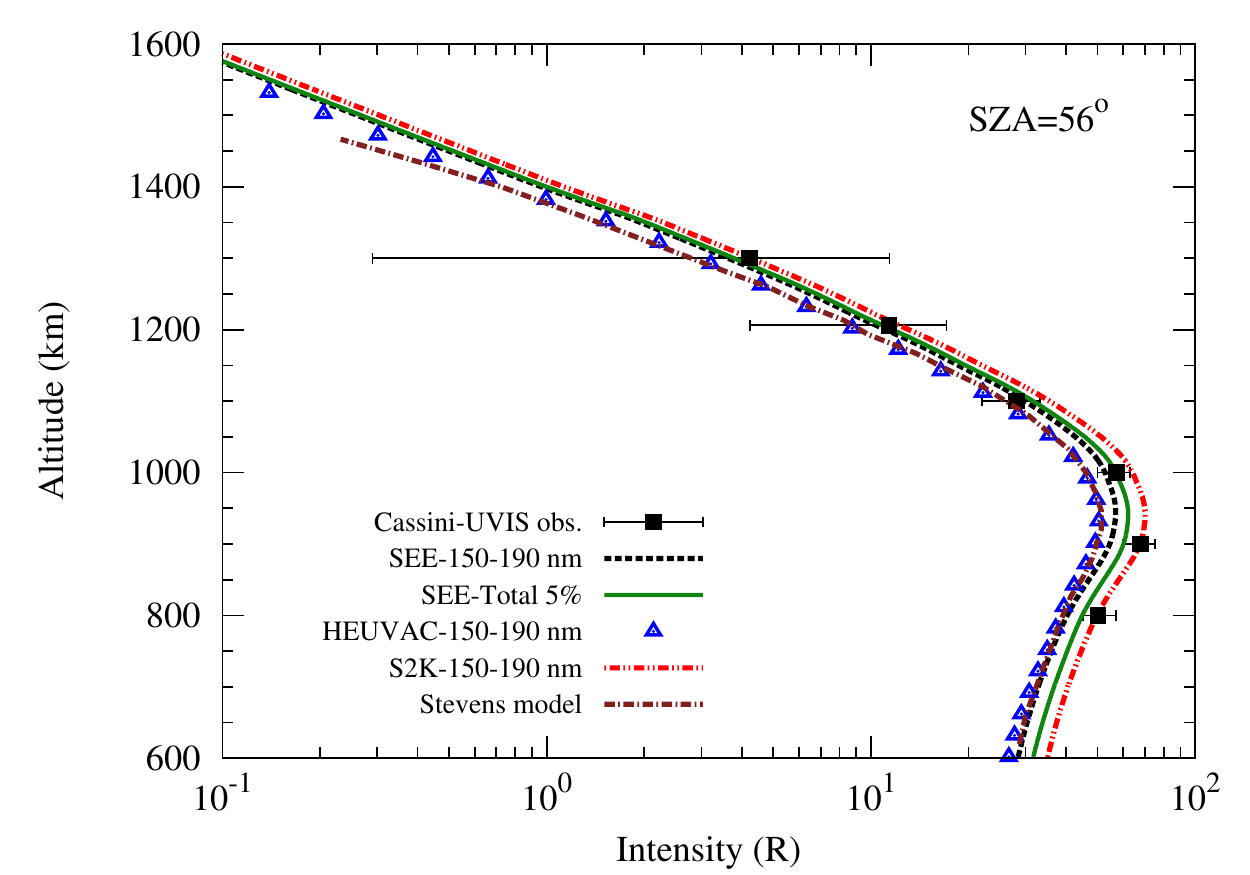}
\caption{Calculated limb profiles of \nt\ VK 150--190 nm wavelength range emissions on 23 June 2009 for
SEE, S2K, and HEUVAC solar EUV flux models at SZA=56\dgr, along with
the Cassini UVIS observed limb intensity and model profile of \cite{Stevens11}. 
Limb profile obtained by taking 5\% of the total \nt\ VK band intensity is also shown.}
\label{fig:int-total}
\end{figure}

As mentioned earlier, \nt\ VK bands were observed for the first time 
in the dayglow of Titan by Cassini UVIS  in the
150--190 nm wavelength band \citep{Stevens11}. For comparing our
calculated limb profiles with UVIS observation we have run
our model at the solar zenith angle of 56\dgr. \cite{Stevens11}
in their calculation assumed that VK bands in the 
150--190 nm range corresponds to  5\% of 
total VK band emission. Figure~\ref{fig:int-total} shows the 
calculated limb intensity of VK bands in the 150--190 nm 
region by taking 5\% of the total VK band intensity, and also
by adding the individual bands which lie in the 150--190 nm 
wavelength region. The calculated limb intensity of
\cite{Stevens11} is also shown in Figure~\ref{fig:int-total} along
with the Cassini--observed limb intensity of VK band in 150--190 nm region
taken from \cite{Stevens11}. We found that VK band emission in 
the wavelength region 150--190 nm corresponds to  4.5\%
of the total VK band intensity (see Table~\ref{tab:oi-triplet}).
Our calculated limb intensity is in good agreement with 
the UVIS observation. The calculated altitude of peak VK emission
also agrees well with the observation within the observational
uncertainty of 15\% \citep{Stevens11}. Our calculated limb intensities
are slightly higher ($ \sim $10\%) than those calculated by \cite{Stevens11}.
Altitude of emission peak is in good agreement in both calculations and is
consistent with that of the observed emission peak. Overall 
good agreement between calculated and observed emission shows that the
VK band intensity can be explained by taking the photoelectron impact
excitation source alone.

\section{Ef{}fect of various model input parameters}
\subsection{Ef{}fect of solar EUV flux model}
\label{subsec:sf-ef{}fect}
To access the ef{}fect of solar EUV flux on the calculated limb intensity we use
the solar EUV flux on 23 June 2009 from SOLAR2000 (S2K) v.2.36 model of
\cite{Tobiska04} and HEUVAC model of \cite{Richards06}. As discussed in 
Section~\ref{sec:mi}, the solar flux at wavelengths below 60 nm
is higher in the S2K model (cf. Figure~\ref{fig:solar-flux}).
The ef{}fect of higher flux at shorter wavelengths is clearly seen in the photoelectron
flux calculations, where the flux calculated using S2K model is higher
than that calculated using SEE solar EUV flux (cf. Figure~\ref{fig:pef}).
While the HEUVAC solar flux is smaller than the SEE solar
flux at wavelengths higher than 15 nm, at wavelengths below 15 nm the
HEUVAC solar flux is higher than SEE flux. Higher EUV fluxes at shorter
wavelength would cause higher photoelectron flux at higher energies ($>$60 eV)
(see Figure.~\ref{fig:pef}).

Volume emission rates of total VK band calculated using S2K and HEUVAC solar
flux models are shown in Figure~\ref{fig:ver-cmp}. The
emission rate calculated using S2K model is $ \sim $23\%  higher than that
calculated using SEE solar EUV flux. At the emission peak and above, volume emission
rate calculated using HEUVAC model is around 14\% smaller than that calculated using 
SEE solar flux. However, below the emission peak, the volume emission rate calculated
using HEUVAC model becomes higher than that calculated using SEE model due
to the larger photoelectron fluxes at higher energies (cf. Figure~\ref{fig:pef}).
At the lower peak, volume emission rate calculated using HEUVAC model is around
2.5 times higher than that calculated using SEE solar EUV flux.

Figures~\ref{fig:int-vk} and \ref{fig:int-total} show the limb intensity of VK band
calculated using S2K and HEUVAC solar flux models. Limb intensities calculated using
S2K solar EUV flux model are slightly higher than the observed values whereas
intensities calculated using SEE and HEUVAC solar fluxes and model predicted intensity of
\cite{Stevens11} are lower than the observed values; but all model calculations are
in good agreement with the observation within the observation and model uncertainties.
The height of peak emission rate (Figure~\ref{fig:ver-cmp}) and altitude of peak
limb intensity (Figures~\ref{fig:int-vk} and \ref{fig:int-total}) are
unaf{}fected by change in input solar EUV flux model.

By changing the solar EUV flux in model, maximum variation at peak intensity is around 40\%.
Since uncertainty in various solar EUV flux itself varies dif{}ferently for dif{}ferent
solar EUV flux, e.g., for SEE observations uncertainty is around 10--20\% \citep{Woods05},
uncertainty in HEUVAC  model depend on the uncertainties in the F74113 reference spectrum and it
could be 15--30\% \citep{Richards06}. Hence, based on the calculations carried out in the present
paper, it is dif{}ficult to suggest a preferred solar flux model. However, since
HEUVAC and S2K depend on various proxies, while SEE solar EUV flux is based on
actual observation and is available online
(\url{http://lasp.colorado.edu/see/l3_data_page.html}) it is preferable to use SEE solar EUV flux.

\subsection{Ef{}fect of solar cycle}
The \nt\ VK band emissions reported by \cite{Stevens11} are for low solar
activity condition (F10.7 = 68, on 23 June 2009). To calculate
the emission intensity during solar maximum we have run our model 
for solar EUV flux on 30 January 2002 (F10.7 = 245) measured by SEE at solar
zenith angle 60\dgr. Model atmosphere remains the same for the solar maximum
calculation. Figure~\ref{fig:ver-cmp} shows the volume emission 
rate of total \nt\ VK band for solar maximum condition. At the 
altitude where emission rate peaks, the calculated emission rate for
solar maximum condition is a factor of $ \sim $2.8 higher than that for
solar minimum condition. Altitude of peak emission rate remains the same 
for both low and high solar activity condition since the model atmosphere
is same. Overhead intensities of individual transition in various
triplet states are also a factor of 2.8 to 2.9 higher for
the solar maximum condition (cf. Table~\ref{tab:oi-triplet}). Similar increase
of around factor of 2.8 can
be seen in limb intensity of total \nt\ VK band for solar maximum 
condition (see Figure~\ref{fig:int-bands}). We have also run our
model for the moderate solar activity condition using the
SEE solar EUV flux for 20 June 2002 (F10.7 = 150) at SZA = 60\dgr.
Model atmosphere remains the same.
Table~\ref{tab:oi-triplet} shows the calculated overhead intensities
of various triplet states during the moderate solar activity condition,
which are a factor of 2 higher than those calculated during
solar minimum condition.

\subsection{Ef{}fect of model atmosphere}
\label{subsec:atm-ef{}fect}
Huygens Atmospheric Structure Instrument (HASI) measured the
density profile of \nt\ in  Titan's atmosphere \citep{Fulchignoni05}. In our
calculations we have used the density profile of \nt\ 
observed by HASI, but reduced by a factor of 3.1 
to be consistent with \nt\ densities measured by 
Ion and Neutral Mass Spectrometer (INMS) \citep{DeLaHaye07}
at 950 km. This reduction is also required for the better
agreement in the UVIS-observed and calculated emission peak 
altitudes. To see the ef{}fect of higher \nt\ density
on emission intensities, we have run our model using the original unscaled
HASI \nt\ density profile keeping the other model input parameters
same. Figure~\ref{fig:int-vk} shows the limb profile of
\nt\ VK band calculated using HASI \nt\ density.
The altitude of the peak emission is situated at
1052 km with a value of $\sim$1.12 kR when HASI \nt\ density
profile is used. Thus, the calculated altitude of peak emission
is around 100 km higher than that calculated by using reduced (by a factor of 3.1)
\nt\ density, but the intensity of emission at the peak remains the
same (1.11 kR).

\section{Summary}
A model for the production of \nt\ triplet band emissions in the dayglow of Titan
is developed. We have used the Analytical Yield Spectra technique to calculate the 
steady state photoelectron flux, which is compared with the
Cassini CAPS observed photoelectron flux. The calculated photoelectron
flux is in good agreement with the observed spectrum in the
6--60 eV energy range. Volume production rates of various
triplet states of \nt\ have been calculated. Population of
any given level of triplet states has been calculated considering 
direct electron impact excitation and quenching as well
as cascading from higher triplet states in statistical
equilibrium condition. Volume emission rates are calculated
and vertically-integrated to calculate the overhead intensities 
of Vegard-Kaplan in various wavelength regions, viz., 130--150, 150--200,
200--300, 300--400, and 400--800 nm, which are given in Table~\ref{tab:oi-triplet}.
In addition, the vertical-integrated intensities of First Positive, Second Positive, Wu-Benesch,
$ B' \rightarrow B $, $ E \rightarrow C$, $ E \rightarrow B $,
$ E \rightarrow A $, and Reverse First Positive bands of \nt, are also calculated and
presented in Table~\ref{tab:oi-triplet} along with their contributions in dif{}ferent
wavelength regions.
Vertically-integrated overhead intensities of various vibrational transitions in triplet
states are presented in Tables~\ref{tab:vk-oi} to \ref{tab:r1p-oi}.

The calculated volume emission rates are integrated along the line of sight
to calculate the limb intensity of total
VK band. Limb profiles of various prominent transitions of
VK band are also calculated and presented in
Figure~\ref{fig:int-bands}. The VK band in the wavelength 
region 200--400 nm contribute around 73\% of the total
VK band intensity, followed by the VK band in
visible region (400--800 nm) which contributes around 22\%.
The calculated limb intensity profile of VK 150--190 nm
band  is in good agreement with the recent Cassini
UVIS-observed profile. We found that the observed intensity
of VK bands can be explained by the photoelectron impact
excitation  alone \citep{Stevens11}. The
ef{}fect of change in solar zenith angle is seen
in the altitude of peak emission as well as intensity at 
the emission peak. Variation in the SZA from 0\dgr\ to 80\dgr\ resulted in
$\sim$18\% upward shift in the  altitude of emission peak, while
the limb intensity at the peak decreased by a factor of 5.5. Our calculation
suggests that intensity of CI 156.1 and 165.7 nm emissions due to photoelectron
impact dissociative excitation of \ch\ and fluorescence scattering of solar lines by 
carbon in Titan's atmosphere are a few orders of magnitude smaller than the
\nt\ VK bands 11--0 (156.3 nm) and 8--0 (165.4 nm) emission intensities, respectively.

We have also made a detailed study on the ef{}fect of solar EUV flux models
on the \nt\ triplet band emission intensities which is a step further to the calculations
of \cite{Stevens11}. Emission intensities
calculated by using the S2K model are around 23\% higher than that calculated using 
the SEE solar flux. The limb intensity at peak calculated using the HEUVAC model
is around 13\% smaller than that calculated using SEE solar flux.
The calculated intensities for moderate (F10.7 = 150) and high (F10.7 = 240) solar activity
conditions are about a factor of 2 and 2.8, respectively, higher
than those calculated during solar minimum (F10.7 = 68)  condition. Calculations
are also carried out taking the HASI-observed \nt\ density in the 
model atmosphere. Due to higher \nt\ density in the HASI observation by a
factor of 3.1, the altitude
of peak emission shifted upwards by 100 km; however, the intensity at the peak remains
the same.

The calculations presented in this paper will help in understanding the production
of \nt\ VK and other triplet band dayglow emissions on Titan as well as 
in other \nt--containing planetary atmospheres.

\section*{Acknowledgement}
\vspace{-0.5cm}
{\small Sonal Kumar Jain was supported by Senior Research Fellowship of ISRO
during the period of this work.}



\newpage

\renewcommand{\thefootnote}{\fnsymbol{footnote}}

\begin{center}
\begin{table*}
\caption{Vertically-integrated overhead intensities of triplet transitions of \nt.}
\small
\begin{tabular*}{\textwidth}{@{\extracolsep{\fill}}lcccccccc}
\hline
\multirow{3}{3cm}{Band} & \multicolumn{6}{c}{Intensity (R)} \\
\cline{2-7}
		& \multicolumn{3}{c}{Min\footnotemark[1]}& & \multicolumn{2}{c}{SEE} \\ 
\cline{2-4} \cline{6-7}
	&  SEE & S2K & HEUVAC & &  Moderate\footnotemark[4] & Max.\footnotemark[2] \\
\hline
Vegard-Kaplan ($A \rightarrow X$)  (137--1155 nm) & 132 & 163 & 121 & & 257 & 371 \\
\hspace{1cm}130--150 nm & 3.7E-2\footnotemark[3] & 4.6E-2 & 3.4E-2 & & 7.3E-2 & 1E-1 \\
\hspace{1cm}150--200 nm & 6 	& 7.5 & 5.5  & & 12 	& 16.7  \\
\hspace{1cm}200--300 nm & 45.7	& 56  & 42.3 & & 89.3	& 129   \\
\hspace{1cm}300--400 nm & 51.2	& 63  & 47.3 & & 100	& 144   \\
\hspace{1cm}400--800 nm & 28.7	& 35  & 26.4 & & 56	& 80.7	\\
\hspace{2cm}400--500 nm & 20.9	& 25  & 19.2 & & 40.9	& 58.9  \\
\hspace{2cm}500--800 nm & 7.8 	& 9.5 & 7.2  & & 15.1	& 21.8  \\
First Positive ($B \rightarrow A$) (263--94129 nm)	& 114 & 141 & 106 &  & 223 & 322   \\
\hspace{1cm} 600--800 nm 	& 55.7 & 69 & 52 & & 109	& 157   \\
Second Positive ($C \rightarrow B$) (268--1140 nm)	& 19 & 24 &  18	& & 37 & 54.5  \\
\hspace{1cm} 300--400 nm	& 17.5 & 22 & 16.6 &  & 34	&  50.3  \\
Wu--Benesch ($W \rightarrow B$) (399--154631 nm)	& 22    & 28 & 21 & & 43.6 & 63.6	\\
$B' \rightarrow B$ (312--37699 nm) 	& 9.5 	& 12    & 9     & & 18.6 & 27.13	\\
$E \rightarrow A$ (207--303 nm)	& 0.25 	& 0.32  & 0.25  & & 0.50	& 0.74  \\
$E \rightarrow B$ (259--483 nm)	& 4E-2  & 5E-2  & 4E-2  & & 8E-2	& 0.12	\\
$E \rightarrow C$ (1113--10127 nm)	& 0.15 	& 0.19  & 0.14  & & 0.3 & 0.43  \\
R1P\footnotemark[5] ($A \rightarrow B$) (739--74175 nm)	& 22 & 27 & 20	& & 43	& 61.6	\\
\hline
\end{tabular*}
\footnotemark[1]{\small Solar minimum condition (F10.7=68).}\\
\footnotemark[4]{\small Solar moderate condition (F10.7=150).}\\
\footnotemark[2]{\small Solar maximum condition (F10.7=256).}\\
\footnotemark[3]{\small 3.7E-2 = $ 3.7 \times 10^{-2}$.} \\
\footnotemark[5]{\small Reverse First Positive.} \\
{\small SEE=Solar EUV Experiment; S2K=SOLAR2000; HEUVAC solar EUV flux models}
\label{tab:oi-triplet}
\end{table*}
\end{center}

\begin{center}
\begin{table*}
\caption{Overhead intensities (in R) of \nt\ Vegard-Kaplan ($ A^3\Sigma_u^+ 
\rightarrow X^1\Sigma^+_g $) band.} \vspace{1pt}
\small
\begin{tabular*}{\textwidth}{@{\extracolsep{\fill}}lccccccccccccccccccccc}
\hline \noalign{\smallskip}
$ \nu'\setminus\nu'' $ & 0 & 1 & 2 & 3 & 4 & 5 & 6 & 7 & 8 & 9 \\
\hline \noalign{\smallskip}
\multirow{1}{*}{0} & 1.4E-5 & 1.9E-2 & 2.5E-1 & 1.2E+0 & 3.2E+0 & 5.6E+0 & 7.1E+0 & 6.8E+0 & 5.0E+0 & 2.9E+0 \\
& (2010) & (2109) & (2216) & (2334) & (2463) & (2605) & (2762) & (2937) & (3133) & (3354) \\
 \noalign{\smallskip}

\multirow{1}{*}{1} & 3.1E-3 & 1.1E-2 & 2.8E-1 & 1.1E+0 & 1.5E+0 & 7.9E-1 & 1.7E-7 & 1.2E+0 & 4.1E+0 & 6.0E+0 \\
 & (1954) & (2047) & (2148) & (2258) & (2379) & (2511) & (2657) & (2819) & (2998) & (3200) \\
 \noalign{\smallskip}

\multirow{1}{*}{2} & 3.2E-2 & 3.0E-3 & 9.1E-2 & 2.8E-1 & 7.4E-2 & 2.0E-1 & 1.3E+0 & 1.7E+0 & 5.3E-1 & 1.2E-1 \\
 & (1901) & (1990) & (2085) & (2189) & (2302) & (2425) & (2561) & (2711) & (2877) & (3062) \\
 \noalign{\smallskip}

\multirow{1}{*}{3} & 1.1E-1 & 6.2E-2 & 2.5E-3 & 1.3E-2 & 6.4E-2 & 4.8E-1 & 4.4E-1 & 8.7E-8 & 7.1E-1 & 1.5E+0 \\
 & (1853) & (1936) & (2027) & (2125) & (2231) & (2347) & (2474) & (2614) & (2768) & (2938) \\
 \noalign{\smallskip}
  
\multirow{1}{*}{4} & 2.5E-1 & 1.6E-1 & 6.7E-3 & 1.9E-3 & 9.0E-2 & 1.3E-1 & 4.1E-3 & 4.1E-1 & 5.4E-1 & 2.4E-2 \\
 & (1808) & (1887) & (1973) & (2065) & (2166) & (2275) & (2394) & (2524) & (2668) & (2826) \\
 \noalign{\smallskip}
  
\multirow{1}{*}{5} & 4.3E-1 & 2.3E-1 & 5.7E-3 & 1.8E-4 & 2.4E-2 & 5.2E-4 & 1.3E-1 & 2.2E-1 & 1.3E-4 & 3.6E-1 \\
 & (1765) & (1841) & (1923) & (2011) & (2106) & (2209) & (2321) & (2443) & (2577) & (2724) \\
 \noalign{\smallskip}
  
\multirow{1}{*}{6} & 6.9E-1 & 2.4E-1 & 9.4E-4 & 7.8E-3 & 5.6E-4 & 1.7E-2 & 9.5E-2 & 4.3E-3 & 1.8E-1 & 2.6E-1 \\
 & (1726) & (1798) & (1876) & (1960) & (2050) & (2147) & (2253) & (2368) & (2494) & (2632) \\
 \noalign{\smallskip}
  
\multirow{1}{*}{7} & 1.1E+0 & 1.9E-1 & 4.2E-2 & 3.2E-2 & 2.5E-4 & 1.5E-2 & 1.5E-2 & 4.8E-2 & 1.7E-1 & 1.1E-3 \\
& (1689) & (1758) & (1833) & (1912) & (1998) & (2090) & (2191) & (2299) & (2418) & (2547) \\
 \noalign{\smallskip}
  
\multirow{1}{*}{8} & 1.0E+0 & 6.2E-2 & 1.1E-1 & 2.7E-2 & 1.2E-3 & 8.0E-4 & 8.5E-4 & 5.0E-2 & 1.4E-2 & 7.8E-2 \\
 & (1655) & (1721) & (1792) & (1868) & (1950) & (2038) & (2133) & (2236) & (2347) & (2469) \\
 \noalign{\smallskip}
  
\multirow{1}{*}{9} & 5.9E-1 & 2.7E-3 & 9.6E-2 & 4.7E-3 & 7.2E-3 & 3.1E-4 & 1.9E-3 & 8.4E-3 & 4.2E-3 & 4.5E-2 \\
 & (1622) & (1686) & (1754) & (1827) & (1905) & (1989) & (2079) & (2177) & (2283) & (2398) \\
 \noalign{\smallskip}
  
\multirow{1}{*}{10} & 2.6E-1 & 3.2E-3 & 5.0E-2 & 1.4E-4 & 7.6E-3 & 6.4E-5 & 1.1E-4 & 1.1E-4 & 5.7E-3 & 4.5E-3 \\
 & (1592) & (1653) & (1718) & (1788) & (1863) & (1943) & (2030) & (2122) & (2223) & (2332) \\
 \noalign{\smallskip}
  
\multirow{1}{*}{11} & 1.2E-1 & 9.3E-3 & 2.0E-2 & 2.7E-3 & 4.1E-3 & 1.7E-4 & 7.8E-5 & 7.4E-5 & 1.5E-3 & 3.8E-5 \\
 & (1563) & (1622) & (1685) & (1752) & (1824) & (1901) & (1983) & (2072) & (2168) & (2271) \\
 \noalign{\smallskip}
  
\multirow{1}{*}{12} & 5.6E-2 & 1.2E-2 & 7.3E-3 & 4.4E-3 & 1.4E-3 & 7.2E-4 & 1.2E-4 & 1.5E-5 & 1.2E-4 & 5.4E-4 \\
 & (1536) & (1593) & (1654) & (1719) & (1788) & (1861) & (1940) & (2025) & (2116) & (2215) \\
 \noalign{\smallskip}
  
\multirow{1}{*}{13} & 2.7E-2 & 1.0E-2 & 1.8E-3 & 4.0E-3 & 1.7E-4 & 8.9E-4 & 1.8E-5 & 1.2E-5 & 3.1E-7 & 2.9E-4 \\
 & (1511) & (1566) & (1625) & (1687) & (1754) & (1824) & (1900) & (1981) & (2069) & (2163) \\
 \noalign{\smallskip}
  
\multirow{1}{*}{14} & 1.3E-2 & 7.9E-3 & 2.1E-4 & 2.8E-3 & 1.6E-5 & 6.6E-4 & 1.4E-5 & 4.6E-5 & 2.7E-6 & 5.3E-5 \\
 & (1487) & (1541) & (1597) & (1658) & (1722) & (1790) & (1863) & (1941) & (2024) & (2114) \\
 \noalign{\smallskip}
  
\multirow{1}{*}{15} & 6.7E-3 & 5.7E-3 & 8.5E-6 & 1.7E-3 & 2.0E-4 & 3.4E-4 & 8.7E-5 & 3.5E-5 & 7.1E-7 & 1.8E-6 \\
 & (1465) & (1517) & (1572) & (1630) & (1692) & (1758) & (1828) & (1903) & (1983) & (2070) \\
 \noalign{\smallskip}
  
\multirow{1}{*}{16} & 3.4E-3 & 3.9E-3 & 1.7E-4 & 8.3E-4 & 3.4E-4 & 1.1E-4 & 1.3E-4 & 8.2E-6 & 9.5E-6 & 3.8E-7 \\
 & (1444) & (1494) & (1548) & (1604) & (1664) & (1728) & (1796) & (1868) & (1945) & (2028) \\
 \noalign{\smallskip}
  
\multirow{1}{*}{17} & 1.8E-3 & 2.6E-3 & 3.3E-4 & 3.4E-4 & 3.6E-4 & 1.5E-5 & 1.3E-4 & 2.2E-7 & 1.5E-5 & 4.7E-8 \\
 & (1425) & (1473) & (1525) & (1580) & (1638) & (1700) & (1765) & (1835) & (1910) & (1990) \\
 \noalign{\smallskip}
  
\multirow{1}{*}{18} & 9.5E-4 & 1.7E-3 & 4.0E-4 & 1.1E-4 & 3.0E-4 & 1.3E-6 & 9.0E-5 & 9.3E-6 & 1.1E-5 & 8.0E-7 \\
 & (1406) & (1454) & (1504) & (1557) & (1614) & (1674) & (1737) & (1805) & (1877) & (1954) \\
 \noalign{\smallskip}
  
\multirow{1}{*}{19} & 5.1E-4 & 1.1E-3 & 4.0E-4 & 2.0E-5 & 2.1E-4 & 1.9E-5 & 4.6E-5 & 2.0E-5 & 4.3E-6 & 3.1E-6 \\
 & (1389) & (1435) & (1484) & (1536) & (1591) & (1649) & (1711) & (1777) & (1846) & (1921) \\
 \noalign{\smallskip}
  
\multirow{1}{*}{20} & 2.8E-4 & 6.9E-4 & 3.6E-4 & 5.0E-8 & 1.4E-4 & 3.8E-5 & 1.8E-5 & 2.5E-5 & 4.1E-7 & 4.5E-6 \\
 & (1373) & (1418) & (1466) & (1517) & (1570) & (1627) & (1687) & (1750) & (1818) & (1890) \\
  
\hline
\end{tabular*}
{\small Values in brackets show the band origin in \AA.} \\
{\small Calculations are made using SEE solar flux (F10.7=68) on 23 June 2009 and at SZA = 60\dgr.}
\label{tab:vk-oi}
\end{table*}
\end{center}

\addtocounter{table}{-1}

\begin{center}
\begin{table*}
\caption{contd.} \vspace{1pt}
\small
\begin{tabular*}{\textwidth}{@{\extracolsep{\fill}}lcccccccccccccccccccccc}
\hline \noalign{\smallskip}
$ \nu'\setminus\nu'' $ & 10 & 11 & 12 & 13 & 14 & 15 &
16 & 17 & 18 & 19  & 20 \\
\hline
\multirow{1}{*}{0} & 1.4E+0 & 5.3E-1 & 1.6E-1 & 4.1E-2 & 8.5E-3 & 1.4E-3 & 1.9E-4 & 2.0E-5 & 1.6E-6 & 1.0E-7 & 4.8E-9 \\
 & (3604) & (3890) & (4221) &  (4606) & (5062) & (5608) & (6274) & (7106) & (8171) & (9584) & (11548) \\ 
 \noalign{\smallskip}

\multirow{1}{*}{1} & 5.7E+0 & 3.8E+0 & 1.9E+0 & 7.5E-1 & 2.3E-1 & 5.6E-2 & 1.1E-2 & 1.7E-3 & 2.0E-4 & 1.8E-5 & 1.3E-6 \\ 
 & (3427) & (3685) & (3980) &  (4321) & (4719) & (5191) & (5757) & (6449) & (7315) & (8427) & (9908)  \\
\noalign{\smallskip}

\multirow{1}{*}{2} & 1.9E+0 & 4.0E+0 & 4.3E+0 & 3.1E+0 & 1.6E+0 & 5.9E-1 & 1.7E-1 & 3.9E-2 & 6.8E-3 & 9.2E-4 & 9.4E-5 \\ 
 & (3270) & (3503) & (3769) &  (4074) & (4426) & (4838) & (5326) & (5913) & (6633) & (7535) & (8697) \\
\noalign{\smallskip}

\multirow{1}{*}{3} & 7.4E-1 & 3.9E-3 & 1.1E+0 & 2.6E+0 & 2.9E+0 & 2.0E+0 & 9.7E-1 & 3.5E-1 & 9.2E-2 & 1.8E-2 & 2.8E-3 \\ 
 & (3129) & (3342) & (3583) &  (3857) & (4171) & (4536) & (4962) & (5468) & (6078) & (6826) & (7767) \\ 
\noalign{\smallskip}

\multirow{1}{*}{4} & 4.5E-1 & 1.2E+0 & 5.9E-1 & 2.0E-3 & 7.7E-1 & 1.8E+0 & 1.9E+0 & 1.2E+0 & 5.3E-1 & 1.7E-1 & 3.9E-2 \\ 
 & (3002) & (3198) & (3418) &  (3666) & (3949) & (4274) & (4650) & (5092) & (5617) & (6250) &  (7030) \\ 
\noalign{\smallskip}

\multirow{1}{*}{5} & 5.1E-1 & 2.0E-2 & 3.9E-1 & 9.2E-1 & 3.7E-1 & 2.0E-2 & 7.0E-1 & 1.3E+0 & 1.2E+0 & 6.8E-1 & 2.6E-1 \\ 
 & (2887) & (3068) & (3270) &  (3497) & (3753) & (4065) & (4381) & (4771) & (5229) & (5773) &  (6432) \\  
\noalign{\smallskip}

\multirow{1}{*}{6} & 3.9E-4 & 4.0E-1 & 4.4E-1 & 1.0E-3 & 4.6E-1 & 7.5E-1 & 1.8E-1 & 8.7E-2 & 7.4E-1 & 1.1E+0 & 8.2E-1 \\ 
 & (2783) & (2951) & (3137) &  (3346) & (3579) & (3844) & (4146) & (4494) & (4897) & (5372) &  (5938) \\  
\noalign{\smallskip}

\multirow{1}{*}{7} & 2.7E-1 & 2.6E-1 & 2.0E-2 & 5.1E-1 & 3.3E-1 & 2.2E-2 & 6.0E-1 & 6.1E-1 & 4.6E-2 & 2.3E-1 & 8.3E-1 \\ 
 & (2689) & (2845) & (3018) &  (3210) & (3424) & (3666) & (3940) & (4252) & (4612) & (5030) &  (5523) \\  
\noalign{\smallskip}

\multirow{1}{*}{8} & 1.4E-1 & 5.2E-3 & 2.6E-1 & 1.2E-1 & 7.2E-2 & 3.9E-1 & 1.1E-1 & 9.3E-2 & 4.6E-1 & 2.5E-1 & 1.2E-3 \\ 
 & (2602) & (2748) & (2909) &  (3087) & (3285) & (3507) & (3757) & (4040) & (4363) & (4736) &  (5170)  \\ 
\noalign{\smallskip}

\multirow{1}{*}{9} & 1.3E-3 & 7.0E-2 & 5.1E-2 & 2.4E-2 & 1.4E-1 & 1.6E-2 & 8.7E-2 & 1.6E-1 & 6.8E-3 & 1.0E-1 & 1.9E-1 \\ 
 & (2523) & (2660) & (2810) &  (2976) & (3160) & (3364) & (3594) & (3852) & (4144) & (4480) &  (4866)  \\ 
\noalign{\smallskip}

\multirow{1}{*}{10} & 8.2E-3 & 2.1E-2 & 1.3E-3 & 4.0E-2 & 7.0E-3 & 2.8E-2 & 4.6E-2 & 2.3E-4 & 5.6E-2 & 3.8E-2 & 2.9E-3 \\ 
 & (2450) & (2579) & (2720) &  (2875) & (3046) & (3236) & (3448) & (3684) & (3952) & (4255) &  (4602)  \\  
\noalign{\smallskip}

\multirow{1}{*}{11} & 5.8E-3 & 5.5E-4 & 8.7E-3 & 5.6E-3 & 5.3E-3 & 1.6E-2 & 1.6E-5 & 2.0E-2 & 8.9E-3 & 5.6E-3 & 2.5E-2 \\ 
 & (2383) & (2505) & (2638) &  (2783) & (2743) & (3120) & (3316) & (3535) & (3780) & (4057) &  (4371)  \\   
\noalign{\smallskip}

\multirow{1}{*}{12} & 1.2E-3 & 7.7E-4 & 3.6E-3 & 1.8E-4 & 6.3E-3 & 4.1E-4 & 6.4E-3 & 4.3E-3 & 2.3E-3 & 9.8E-3 & 3.7E-4 \\ 
 & (2321) & (2437) & (2562) &  (2699) & (2850) & (3015) & (3198) & (3400) & (3627) & (3881) &  (4168)  \\  
\noalign{\smallskip}

\multirow{1}{*}{13} & 2.6E-5 & 1.0E-3 & 2.6E-4 & 1.5E-3 & 1.1E-3 & 1.2E-3 & 2.7E-3 & 2.7E-4 & 4.1E-3 & 2.6E-4 & 3.3E-3 \\ 
 & (2264) & (2374) & (2493) &  (2622) & (2764) & (2919) & (3090) & (3279) & (3489) & (3724) &  (3987)  \\   
\noalign{\smallskip}

\multirow{1}{*}{14} & 4.2E-5 & 3.5E-4 & 6.1E-5 & 8.2E-4 & 1.4E-5 & 1.3E-3 & 4.3E-5 & 1.5E-3 & 4.9E-4 & 1.1E-3 & 1.5E-3 \\ 
 & (2211) & (2316) & (2429) &  (2552) & (2686) & (2832) & (2993) & (3169) & (3365) & (3583) &  (3826)  \\   
\noalign{\smallskip}

\multirow{1}{*}{15} & 5.5E-5 & 4.0E-5 & 1.9E-4 & 1.3E-4 & 3.0E-4 & 3.1E-4 & 2.9E-4 & 6.0E-4 & 1.4E-4 & 9.4E-4 & 2.2E-6 \\ 
 & (2162) & (2262) & (2370) &  (2487) & (2614) & (2752) & (2904) & (3070) & (3253) & (3456) &  (3682)  \\   
\noalign{\smallskip}

\multirow{1}{*}{16} & 2.0E-5 & 4.0E-7 & 1.1E-4 & 5.3E-7 & 2.3E-4 & 2.2E-7 & 3.5E-4 & 1.3E-5 & 4.2E-4 & 8.0E-5 & 3.8E-4 \\ 
 & (2117) & (2213) & (2316) &  (2427) & (2548) & (2679) & (2823) & (2980) & (3152) & (3342) &  (3553)  \\   
\noalign{\smallskip}

\multirow{1}{*}{17} & 2.5E-6 & 8.0E-6 & 2.8E-5 & 3.1E-5 & 6.5E-5 & 5.6E-5 & 1.1E-4 & 6.5E-5 & 1.7E-4 & 4.8E-5 & 2.5E-4 \\ 
 & (2075) & (2167) & (2266) &  (2372) & (2488) & (2613) & (2749) & (2897) & (3060) & (3239) &  (3436)  \\   
\noalign{\smallskip}

\multirow{1}{*}{18} & 2.1E-10 & 6.2E-6 & 1.9E-6 & 3.3E-5 & 3.3E-6 & 6.8E-5 & 4.8E-6 & 1.0E-4 & 9.4E-6 & 1.3E-4 & 2.2E-5\\ 
 & (2036) & (2125) & (2220) &  (2322) & (2432) & (2551) & (2681) & (2822) & (2976) & (3145) &  (3331)  \\ 
\noalign{\smallskip}

\multirow{1}{*}{19} & 1.3E-7 & 1.8E-6 & 3.6E-7 & 1.5E-5 & 2.8E-6 & 3.2E-5 & 7.7E-6 & 4.8E-5 & 1.2E-5 & 9.4E-5 & 1.2E-5 \\ 
 & (2000) & (2086) & (2177) &  (2275) & (2381) & (2495) & (2619) & (2754) & (2900) & (3060) &  (3236)  \\   
\noalign{\smallskip}

\multirow{1}{*}{20} & 4.0E-9 & 1.5E-7 & 1.3E-6 & 3.6E-6 & 7.8E-6 & 6.4E-6 & 1.9E-5 & 7.5E-6 & 3.0E-5 & 8.6E-6 & 4.0E-5 \\ 
 & (1967) & (2050) & (2138) &  (2232) & (2334) & (2444) & (2563) & (2691) & (2831) & (2984) &  (3150)   \\ 
\hline
\end{tabular*}
\end{table*}
\end{center}

\begin{center}
\begin{table*}
\caption{Overhead intensities (in R) of \nt\ First Positive ($ B^3\Pi_g
\rightarrow A^3\Sigma^+_u $) band.} \vspace{1pt}
\small
\begin{tabular*}{\textwidth}{@{\extracolsep{\fill}}lccccccccccccccccccccc}
\hline \noalign{\smallskip}
$ \nu'\setminus\nu'' $ & 0 & 1 & 2 & 3 & 4 & 5 & 6 & 7 & 8 & 9 \\
\hline \noalign{\smallskip}
\multirow{1}{*}{0} & 1.1E+1 & 5.7E+0 & 1.7E+0 & 3.6E-1 & 5.6E-2 & 5.8E-3 & 2.6E-4 & 3.9E-8 & -- & --  \\
& (10469) & (12317) & (14895) & (18739) & (25084) & (37523) & (72916) & (941292) & -- & -- \\
 \noalign{\smallskip}

\multirow{1}{*}{1} & 1.5E+1 & 6.7E-2 & 2.8E+0 & 2.1E+0 & 7.8E-1 & 1.9E-1 & 2.9E-2 & 2.4E-3 & 3.0E-5 & -- \\
 & (8883) & (10179) & (11878) & (14201) & (17569) & (22882) & (32502) & (55202) & (173933) & -- \\
 \noalign{\smallskip}

\multirow{1}{*}{2} & 6.9E+0 & 9.0E+0 & 1.7E+0 & 3.4E-1 & 1.3E+0 & 8.4E-1 & 3.0E-1 & 6.8E-2 & 9.2E-3 & 4.1E-4 \\
 & (7732) & (8695) & (9905) & (11471) & (13572) & (16538) & (21039) & (28671) & (44420) & (95828) \\
 \noalign{\smallskip}

\multirow{1}{*}{3} & 1.4E+0 & 9.3E+0 & 2.4E+0 & 2.9E+0 & 7.6E-2 & 3.6E-1 & 5.6E-1 & 3.1E-1 & 1.0E-1 & 2.0E-2 \\
 & (6858) & (7606) & (8516) & (9648) & (11092) & (12997) & (15624) & (19474) & (25651) & (37163) \\
 \noalign{\smallskip}
  
\multirow{1}{*}{4} & 1.4E-1 & 2.8E+0 & 7.1E+0 & 1.3E-1 & 2.1E+0 & 5.3E-1 & 1.3E-2 & 2.3E-1 & 2.2E-1 & 1.0E-1 \\
 & (6173) & (6772) & (7484) & (8345) & (9404) & (10739) & (12471) & (14807) & (18126) & (23207) \\
 \noalign{\smallskip}
  
\multirow{1}{*}{5} & 6.4E-3 & 3.5E-1 & 3.2E+0 & 3.8E+0 & 1.4E-1 & 8.9E-1 & 6.7E-1 & 4.5E-2 & 4.1E-2 & 1.1E-1 \\
 & (5622) & (6114) & (6689) & (7368) & (8181) & (9173) & (10408) & (11987) & (14072) & (16954) \\
 \noalign{\smallskip}
  
\multirow{1}{*}{6} & 1.3E-4 & 2.0E-2 & 5.0E-1 & 2.6E+0 & 1.5E+0 & 4.7E-1 & 2.0E-1 & 4.6E-1 & 1.3E-1 & 2.1E-5 \\
 & (5168) & (5582) & (6057) & (6608) & (7255) & (8025) & (8954) & (10098) & (11539) & (13407) \\
 \noalign{\smallskip}
  
\multirow{1}{*}{7} & 1.0E-6 & 4.8E-4 & 3.4E-2 & 5.3E-1 & 1.7E+0 & 4.0E-1 & 4.9E-1 & 9.0E-3 & 2.2E-1 & 1.3E-1 \\
& (4789) & (5142) & (5543) & (6001) & (6530) & (7147) & (7875) & (8746) & (9806) & (11124) \\
 \noalign{\smallskip}
  
\multirow{1}{*}{8} & 1.0E-9 & 3.9E-6 & 9.2E-4 & 4.1E-2 & 4.4E-1 & 9.6E-1 & 5.5E-2 & 3.3E-1 & 1.1E-2 & 6.7E-2 \\
 & (4448) & (4774) & (5117) & (5505) & (5947) & (6454) & (7042) & (7730) & (8548) & (9532) \\
 \noalign{\smallskip}
  
\multirow{1}{*}{9} & 3.0E-12 & 4.0E-9 & 8.5E-6 & 1.3E-3 & 4.3E-2 & 3.3E-1 & 4.8E-1 & 5.3E-5 & 1.7E-1 & 3.6E-2 \\
 & (4192) & (4460) & (4758) & (5092) & (5468) & (5894) & (6380) & (6940) & (7592) & (8358) \\
 \noalign{\smallskip}
  
\multirow{1}{*}{10} & 1.3E-13 & 3.5E-11 & 8.0E-9 & 1.3E-5 & 1.5E-3 & 3.7E-2 & 2.2E-1 & 2.0E-1 & 1.1E-2 & 6.7E-2 \\
 & (3953) & (4190) & (4453) & (4744) & (5068) & (5432) & (5842) & (6309) & (6843) & (7459) \\
 \noalign{\smallskip}
  
\multirow{1}{*}{11} & 7.4E-15 & 5.6E-14 & 7.8E-11 & 1.1E-8 & 1.5E-5 & 1.4E-3 & 2.7E-2 & 1.2E-1 & 7.2E-2 & 1.8E-2 \\
 & (3744) & (3956) & (4189) & (4446) & (4729) & (5045) & (5397) & (5792) & (6239) & (6748) \\
 \noalign{\smallskip}
  
\multirow{1}{*}{12} & 8.4E-14 & 7.7E-15 & 8.0E-13 & 1.3E-10 & 1.1E-8 & 1.4E-5 & 1.1E-3 & 1.7E-2 & 5.8E-2 & 1.9E-2 \\
 & (3560) & (3751) & (3960) & (4188) & (4439) & (4716) & (5022) & (5363) & (5744) & (6172) \\
 \noalign{\smallskip}
  
\multirow{1}{*}{13} & 2.1E-14 & 3.0E-14 & 1.0E-13 & 3.3E-12 & 2.0E-10 & 9.2E-9 & 1.1E-5 & 7.3E-4 & 9.6E-3 & 2.6E-2 \\
 & (3396) & (3570) & (3758) & (3963) & (4187) & (4433) & (4702) & (5000) & (5329) & (5696) \\
 \noalign{\smallskip}
  
\multirow{1}{*}{14} & 4.4E-16 & 4.2E-14 & 9.2E-14 & 7.2E-13 & 8.5E-12 & 2.8E-10 & 7.7E-9  & 9.2E-6  & 5.3E-4  & 5.8E-3 \\
 & (3250) & (3409) & (3580) & (3766) & (3968) & (4187) & (4427) & (4690) & (4978) & (5297) \\
 \noalign{\smallskip}
  
\multirow{1}{*}{15} & 5.5E-15 & 6.1E-15 & 6.9E-14 & 3.7E-13 & 1.7E-12 & 1.6E-11 & 3.6E-10 & 6.7E-9 & 8.2E-6 & 4.2E-4 \\
 & (3119) & (3265) & (3422) & (3591) & (3774) & (3972) & (4188) & (4422) & (4678) & (4958) \\
 \noalign{\smallskip}
  
\multirow{1}{*}{16} & 1.2E-15 & 1.4E-16 & 3.0E-14 & 1.4E-13 & 6.7E-13 & 3.1E-12 & 2.9E-11 & 4.2E-10 & 5.5E-9 & 6.9E-6 \\
 & (3001) & (3136) & (3280) & (3436) & (3603) & (3783) & (3978) & (4188) & (4417) & (4666) \\
 \noalign{\smallskip}
  
\multirow{1}{*}{17} & 2.4E-16 & 6.2E-17 & 5.1E-15 & 2.9E-14 & 1.8E-13 & 9.8E-13 & 4.0E-12 & 3.3E-11 & 3.4E-10 & 3.3E-9 \\
 & (2894) & (3020) & (3153) & (3296) & (3450) & (3615) & (3792) & (3983) & (4190) & (4413) \\
 \noalign{\smallskip}
  
\multirow{1}{*}{18} & 8.0E-16 & 3.0E-16 & 2.1E-16 & 4.0E-15 & 3.9E-14 & 2.3E-13 & 9.6E-13 & 3.9E-12 & 3.1E-11 & 2.3E-10 \\
 & (2797) & (2914) & (3039) & (3171) & (3313) & (3465) & (3627) & (3802) & (3989) & (4191) \\
 \noalign{\smallskip}
  
\multirow{1}{*}{19} & 3.9E-16 & 5.6E-16 & 2.5E-17 & 4.7E-16 & 9.4E-15 & 5.5E-14 & 3.0E-13 & 1.2E-12 & 4.8E-12 & 3.8E-11 \\
 & (3709) & (2818) & (2935) & (3058) & (3190) & (3330) & (3480) & (3641) & (3812) & (3996) \\
 \noalign{\smallskip}
  
\multirow{1}{*}{20} & 5.7E-18 & 7.3E-17 & 4.4E-17 & 1.3E-17 & 1.0E-15 & 7.5E-15 & 5.2E-14 & 2.5E-13 & 9.6E-13 & 4.2E-12 \\
 & (2628) & (2731) & (2840) & (2956) & (3079) & (3209) & (3348) & (3496) & (3654) & (3823) \\
  
\hline
\end{tabular*}
\\
{\small Values in brackets show the band origin in \AA.} \\
{\small Calculations are made using SEE solar flux on 23 June 2009 and at SZA = 60\dgr.}
\label{tab:BA-oi}
\end{table*}
\end{center}

\addtocounter{table}{-1}

\begin{center}
\begin{table*}
\caption{contd.} \vspace{1pt}
\small
\begin{tabular}{lccccccccccccccccccccc}
\hline \noalign{\smallskip}
$ \nu'\setminus\nu'' $ & 10 & 11 & 12 & 13 & 14 & 15 &
16 & 17 & 18 & 19  & 20 \\
\hline

\multirow{1}{*}{3}  & 2.0E-2 & 1.9E-3 & -- & -- & -- & -- & -- & -- & -- & -- & --  \\
 & (66117) & (274786) & -- & -- & -- & -- & -- & -- & -- & -- & --  \\
\noalign{\smallskip}

\multirow{1}{*}{4} & 3.0E-2 & 4.8E-3 & 2.3E-4 & -- & -- & -- & -- & -- & -- & -- & --  \\
 & (31941) & (50449) & (115737) & -- & -- & -- & -- & -- & -- & -- & --  \\
\noalign{\smallskip}

\multirow{1}{*}{5} & 7.8E-2 & 3.2E-2 & 7.7E-3 & 8.7E-4 & 6.2E-6 & -- & -- & -- & -- & -- &  -- \\ 
 & (21186) & (27999) & (40761) &  (73199) & (311787) & -- & -- & -- & -- & -- &  -- \\  
\noalign{\smallskip}

\multirow{1}{*}{6} & 3.1E-2 & 4.2E-2 & 2.5E-2 & 8.6E-3 & 1.7E-3 & 1.1E-4 &-- & -- & -- & -- &  -- \\ 
 & (15923) & (19487) & (24916) &  (34172) & (53448) & (117886) &-- & -- & -- & -- &  -- \\
\noalign{\smallskip}

\multirow{1}{*}{7} & 1.5E-2 & 2.9E-3 & 1.5E-2 & 1.5E-2 & 7.4E-3 & 2.3E-3 & 3.5E-4 & 7.7E-6 & -- & -- &  -- \\
 & (12802) & (15009) & (18036) &  (22434) & (29394) & (42028) & (71926) & (229296) & -- & -- &  -- \\  
\noalign{\smallskip}

\multirow{1}{*}{8} & 8.9E-2 & 2.8E-2 & 7.5E-4 & 3.0E-3 & 6.3E-3 & 4.8E-3 & 2.1E-3 & 5.8E-4 & 6.7E-5 & 1.2E-7 & -- \\ 
 & (10737) & (12248) & (14192) &  (16781) & (20392) & (25765) & (34576) & (51601) & (98056) & (718231) & --  \\ 
\noalign{\smallskip}

\multirow{1}{*}{9} & 1.1E-2 & 4.4E-2 & 2.6E-2 & 4.8E-3 & 2.3E-5 & 1.7E-3 & 2.4E-3 & 1.6E-3 & 6.5E-4 & 1.6E-4 & 1.4E-5 \\
 & (9272) & (10377) & (11739) &  (13456) & (15683) & (18679) & (22912) & (29322) & (40124) & (62048) &  (129839)  \\ 
\noalign{\smallskip}

\multirow{1}{*}{10} & 3.8E-2 & 4.6E-5 & 1.5E-2 & 1.7E-2 & 6.4E-3 & 6.3E-4 & 1.4E-4 & 7.7E-4 & 8.4E-4 & 5.2E-4 & 2.1E-4  \\ 
 & (8178) & (9025) & (10038) &  (11268) & (12789) & (14713) & (17219) & (20605) & (25412) & (32758) &  (45186)  \\  
\noalign{\smallskip}

\multirow{1}{*}{11} & 1.9E-2 & 2.5E-2 & 1.8E-3 & 3.1E-3 & 7.6E-3 & 4.9E-3 & 1.3E-3 & 5.4E-5 & 1.1E-4 & 3.0E-4 & 2.9E-4 \\
 & (7330) & (8004) & (8791) &  (9720) & (10831) & (12181) & (13849) & (15958) & (18697) & (22383) &  (27577)  \\
\noalign{\smallskip}

\multirow{1}{*}{12} & 1.5E-2 & 1.8E-3 & 1.2E-2 & 3.1E-3 & 1.5E-4 & 2.3E-3 & 2.5E-3 & 1.2E-3 & 2.4E-4 & 1.4E-6 & 5.0E-5 \\
 & (6656) & (7207) & (7838) &  (8568) & (9420) & (10424) & (11623) & (13072) & (14855) & (17091) &  (19962)  \\  
\noalign{\smallskip}

\multirow{1}{*}{13} & 4.2E-3 & 8.6E-3 & 1.1E-4 & 4.3E-3 & 2.5E-3 & 8.3E-5 & 4.3E-4 & 9.5E-4 & 7.2E-4 & 2.8E-4 & 4.6E-5 \\
 & (6106) & (6566) & (7087) &  (7678) & (8355) & (9136) & (10043) & (11108) & (12369) & (13881) &  (15717)  \\   
\noalign{\smallskip}

\multirow{1}{*}{14} & 1.2E-2 & 6.4E-4 & 4.7E-3 & 9.9E-5 & 1.4E-3 & 1.6E-3 & 3.1E-4 & 2.5E-5  & 2.9E-4 & 3.6E-4 & 2.2E-4 \\
 & (5650) & (6042) & (6479) &  (6970) & (7524) & (8151) & (8866) & (9685) & (10630) & (11728) &  (13012)  \\
\noalign{\smallskip}

\multirow{1}{*}{15} & 3.9E-3 & 6.2E-3 & 1.8E-5 & 2.5E-3 & 3.5E-4 & 3.6E-4 & 9.0E-4 & 3.9E-4 & 1.5E-5 & 5.8E-5 & 1.5E-4 \\
 & (5265) & (5604) & (5979) &  (6394) & (6857) & (7374) & (7955) & (8608) & (9347) & (10185) &  (11140)  \\
\noalign{\smallskip}

\multirow{1}{*}{16} & 3.2E-4 & 2.5E-3 & 3.0E-3 & 4.9E-5 & 1.2E-3 & 4.4E-4 & 4.4E-5 & 4.1E-4 & 3.2E-4 & 6.9E-5 & 9.3E-7 \\
 & (4938) & (5234) & (5560) &  (5917) & (6312) & (6747) & (7230) & (7765) & (8312) & (9026) &  (9768)  \\ 
\noalign{\smallskip}

\multirow{1}{*}{17} & 4.4E-6 & 1.8E-4 & 1.2E-3 & 1.1E-3 & 1.1E-4 & 3.6E-4 & 2.9E-4 & 7.5E-7 & 1.1E-4 & 1.6E-4 & 7.1E-5 \\
 & (4655) & (4918) & (5204) &  (5516) & (5857) & (6230) & (6639) & (7089) & (7582) & (8124) &  (8721)  \\
\noalign{\smallskip}

\multirow{1}{*}{18} & 1.9E-9 & 2.6E-6 & 9.8E-5 & 5.5E-4 & 3.4E-4 & 1.0E-4 & 8.6E-5 & 1.5E-4 & 1.4E-5 & 2.0E-5 & 6.0E-5 \\
 & (4409) & (4644) & (4899) &  (5174) & (5473) & (5798) & (6150) & (6534) & (6951) & (7404) &  (7896)  \\ 
\noalign{\smallskip}

\multirow{1}{*}{19} & 2.1E-10 & 1.5E-9 & 2.2E-6 & 7.4E-5 & 3.5E-4 & 1.4E-4 & 1.0E-4 & 2.0E-5 & 8.9E-5 & 2.6E-5 & 1.5E-6 \\
 & (4194) & (4406) & (4634) &  (4880) & (5145) & (5431) & (5739) & (6072) & (6430) & (6816) &  (7231)  \\   
\noalign{\smallskip}

\multirow{1}{*}{20} & 3.3E-11 & 1.4E-10 & 9.7E-10 & 1.4E-6 & 4.3E-5 & 1.7E-4 & 4.2E-5 & 6.3E-5 & 1.3E-6 & 3.6E-5 & 2.1E-5 \\
 & (4003) & (4196) & (4403) &  (4624) & (4862) & (5116) & (5389) & (5681) & (5993) & (6327) &  (6683)   \\ 
\hline
\end{tabular}
\end{table*}
\end{center}

\begin{center}
\begin{table*}
\caption{Overhead intensities (in R) of \nt\ Second Positive ($ C^3\Pi_u \rightarrow B^3\Pi_g$) band.} \vspace{1pt}
\small
\begin{tabular*}{\textwidth}{@{\extracolsep{\fill}}lccccccccccccccccccccc}
\hline \noalign{\smallskip}
$ \nu'\setminus\nu'' $ & 0 & 1 & 2 & 3 & 4  & 5 & 6 & 7 & 8 & 9 \\
\hline \noalign{\smallskip}

\multirow{1}{*}{0} & 6.5E+0 & 4.4E+0 & 1.8E+0 & 5.5E-1 & 1.5E-1 & 3.5E-2 & 7.7E-3 & 1.6E-3 & 3.3E-4 & 6.7E-5 \\
& (3370) & (3576) & (3804) & (4058) & (4343) & (4665) & (5032) & (5452) & (5938) & (6507) \\
 \noalign{\smallskip}

\multirow{1}{*}{1} & 1.7E+0 & 8.3E-2 & 7.8E-1 & 7.0E-1 & 3.4E-1 & 1.3E-1 & 3.9E-2 & 1.1E-2 & 2.8E-3 & 6.6E-4 \\
 & (3158) & (3338) & (3536) & (3754) & (3997) & (4268) & (4573) & (4317) & (5309) & (5759) \\
 \noalign{\smallskip}

\multirow{1}{*}{2} & 2.0E-1 & 5.0E-1 & 3.9E-2 & 8.4E-2 & 2.0E-1 & 1.5E-1 & 7.7E-2 & 3.0E-2 & 1.0E-2 & 3.0E-3 \\
 & (2976) & (3135) & (3309) & (3499) & (3709) & (3942) & (4200) & (4489) & (4813) & (5180) \\
 \noalign{\smallskip}

\multirow{1}{*}{3} & 7.5E-3 & 1.0E-1 & 8.5E-2 & 4.1E-2 & 1.6E-3 & 3.3E-2 & 4.3E-2 & 2.9E-2 & 1.4E-2 & 5.6E-3 \\
 & (2818) & (2961) & (3115) & (3284) & (3468) & (3671) & (3894) & (4140) & (4415) & (4722) \\
 \noalign{\smallskip}
  
\multirow{1}{*}{4} & 5.2E-5 & 4.9E-3 & 3.4E-2 & 1.1E-2 & 1.4E-2 & 4.7E-4 & 3.8E-3 & 8.9E-3 & 7.9E-3 & 4.7E-3 \\
 & (2684) & (2812) & (2952) & (3102) & (3266) & (3445) & (3641) & (3856) & (4093) & (4355) \\
 \noalign{\smallskip}

\hline
\end{tabular*}
\\
{\small Values in brackets show the band origin in \AA.} \\
{\small Calculations are made using SEE solar flux on 23 June 2009 and at SZA = 60\dgr.}
\label{tab:CB-oi}
\end{table*}
\end{center}

\addtocounter{table}{-1}

\begin{center}
\begin{table*}
\caption{contd.} \vspace{1pt}
\small
\begin{tabular}{lccccccccccccccccccccc}
\hline \noalign{\smallskip}
$ \nu'\setminus\nu'' $ & 10 & 11 & 12 & 13 & 14 & 15 &
16 & 17 & 18 & 19  & 20 \\
\hline
\multirow{1}{*}{0} & 1.3E-5 & 2.4E-6 & 4.2E-7 & 6.9E-8 & 1.0E-8 & 1.3E-9 & 1.3E-10 & 7.4E-12 & 5.7E-16 & 2.6E-13 & 7.6E-14 \\
 & (7181) & (7992) & (8985) &  (10228) & (11828) & (13962) & (16944) & (21403) & (28779) & (43302) & (84991) \\ 
 \noalign{\smallskip}

\multirow{1}{*}{1} & 1.5E-4 & 3.3E-5 & 7.1E-6 & 1.4E-6 & 2.7E-7 & 4.6E-14 & 6.8E-09 & 7.6E-10 & 4.4E-11 & 2.3E-14 & 2.2E-12 \\ 
 & (6281) & (6893) & (7619) &  (8495) & (9571) & (10921) & (12665) & (15000) & (18285) & (23236) & (31536)  \\
\noalign{\smallskip}

\multirow{1}{*}{2} & 8.5E-4 & 2.2E-4 & 5.5E-5 & 1.3E-5 & 3.0E-6 & 6.3E-7 & 1.2E-7 & 2.1E-8 & 2.9E-9 & 2.4E-10 & 1.8E-12 \\ 
 & (5599) & (6080) & (6638) &  (7293) & (8071) & (9011) & (10166) & (11618) & (13495) & (16014) & (19563) \\
\noalign{\smallskip}

\multirow{1}{*}{3} & 1.9E-3 & 6.2E-4 & 1.8E-4 & 5.1E-5 & 1.4E-5 & 3.4E-6 & 8.1E-7 & 1.8E-7 & 3.5E-8 & 5.7E-9 & 7.1E-10 \\ 
 & (5067) & (5458) & (5904) &  (6416) & (7011) & (7709) & (8539) & (9541) & (10771) & (12317) & (14315) \\ 
\noalign{\smallskip}

\multirow{1}{*}{4} & 2.2E-3 & 8.8E-4 & 3.2E-4 & 1.0E-4 & 3.2E-5 & 9.4E-6 & 2.6E-6 & 7.0E-7 & 1.7E-7 & 3.7E-8 & 7.1E-9 \\ 
 & (4648) & (4974) & (5342) &  (5758) & (6233) & (6778) & (7412) & (8155) & (9037) & (10101) &  (11406) \\ 
\noalign{\smallskip}
\hline
\end{tabular}
\\
{\small Values in brackets show the band origin in \AA.} \\
{\small Calculations are made using SEE solar flux on 23 June 2009 and at SZA = 60\dgr.}
\end{table*}
\end{center}

\begin{center}
\begin{table*}
\caption{Overhead intensities (in R) of \nt\ Wu-Benesch ($ W^3\Delta_u 
\rightarrow B^3\Pi_g $) band.} \vspace{1pt}
\small
\begin{tabular}{lccccccccccccccccccccc}
\hline \noalign{\smallskip}
$ \nu'\setminus\nu'' $ & 0 & 1 & 2 & 3 & 4 & 5 & 6 & 7 & 8 & 9 \\
\hline \noalign{\smallskip}
\multirow{1}{*}{0} & 1.6E-1 & -- & -- & -- & -- & -- & -- & -- & -- & --  \\
& (1361044) & -- & -- & -- & -- & -- & -- & -- & -- & -- \\
 \noalign{\smallskip}

\multirow{1}{*}{1} & 4.5E-1 & -- & -- & -- & -- & -- & -- & -- & -- & -- \\
 & (64311) & -- & -- & -- & -- & -- & -- & -- & -- & -- \\
 \noalign{\smallskip}

\multirow{1}{*}{2} & 6.5E-1 & 7.3E-2 & -- & -- & -- & -- & -- & -- & -- & -- \\
 & (33206) & (76556) & -- & -- & -- & -- & -- & -- & -- & -- \\
 \noalign{\smallskip}

\multirow{1}{*}{3} & 5.6E-1 & 4.9E-1 & 9.0E-3 & & -- & -- & -- & -- & -- & -- \\
 & (22505) & (36522) & (94180) & -- & -- & -- & -- & -- & -- & -- \\
 \noalign{\smallskip}
  
\multirow{1}{*}{4} & 3.5E-1 & 8.5E-1 & 2.4E-1 & 1.9E-4 & -- & -- & -- & -- & -- & --  \\
 & (17092) & (24124) & (40502) & (121693) & -- & -- & -- & -- & -- & --  \\
 \noalign{\smallskip}
  
\multirow{1}{*}{5} & 1.9E-1 & 7.9E-1 & 7.5E-1 & 8.8E-2 & 1.8E-4 & -- & -- & -- & -- & -- \\
 & (13825) & (18090) & (25962) & (45362) & (170594) & -- & -- & -- & -- & --  \\
 \noalign{\smallskip}
  
\multirow{1}{*}{6} & 8.0E-2 & 5.2E-1 & 9.8E-1 & 4.9E-1 & 1.8E-2 & 1.6E-4 &  -- & -- & -- & -- \\
 & (11639) & (14521) & (19193) & (28067) & (51426) & (281484) & -- & -- & -- & -- \\
 \noalign{\smallskip}
  
\multirow{1}{*}{7} & 3.0E-2 & 2.7E-1 & 7.9E-1 & 8.6E-1 & 2.3E-1 & 8.7E-4 & 1.1E-5 & -- & -- & --  \\
& (10075) & (12165) & (15281) & (20421) & (30501) & (59196) & (774629) & -- & -- & -- \\
 \noalign{\smallskip}
  
\multirow{1}{*}{8} & 1.0E-2 & 1.2E-1 & 4.8E-1 & 8.7E-1 & 5.9E-1 & 8.3E-2 & 5.4E-4 & -- & -- & --   \\
 & (8900) & (10493) & (12732) & (16112) & (21794) & (33343) & (69497) & -- & -- & --  \\
 \noalign{\smallskip}
  
\multirow{1}{*}{9} & 3.3E-3 & 4.5E-2 & 2.4E-1 & 6.2E-1 & 7.5E-1 & 3.3E-1 & 1.9E-2 & 1.5E-3  & -- & -- \\
 & (7986) & (9246) & (10941) & (13347) & (17024) & (23339) & (36704) & (83791) & -- & -- \\
 \noalign{\smallskip}
  
\multirow{1}{*}{10} & 9.8E-4 & 1.6E-2 & 1.0E-1 & 3.5E-1 & 6.2E-1 & 5.3E-1 & 1.5E-1 & 1.6E-3 & 1.2E-3 & -- \\
 & (7255) & (8280) & (9614) & (11424) & (14014) & (18030) & (25088) & (40735) & (104922) & -- \\
 \noalign{\smallskip}
  
\multirow{1}{*}{11} & 2.8E-4 & 5.2E-3 & 4.0E-2 & 1.7E-1 & 4.0E-1 & 5.2E-1 & 3.1E-1 & 5.1E-2 & 2.1E-4 & 5.4E-4 \\
 & (6658) & (7511) & (8592) & (10009) & (11943) & (14742) & (19145) & (27084) & (45654) & (139282) \\
 \noalign{\smallskip}
  
\multirow{1}{*}{12} & 7.9E-5 & 1.6E-3 & 1.5E-2 & 7.2E-2 & 2.1E-1 & 3.8E-1 & 3.7E-1 & 1.6E-1 & 1.2E-2 & 1.3E-3 \\
 & (6160) & (6883) & (7781) & (8925) & (10432) & (12505) & (15536) & (20386) & (29381) & (51783) \\
 \noalign{\smallskip}
  
\multirow{1}{*}{13} & 2.2E-5 & 4.9E-4 & 4.9E-3 & 2.8E-2 & 1.0E-1 & 2.3E-1 & 3.1E-1 & 2.3E-1 & 6.7E-2 & 1.4E-3 \\
 & (5740) & (6363) & (7122) & (8069) & (9281) & (10886) & (13114) & (16408) & (21774) & (32050) \\
 \noalign{\smallskip}
  
\multirow{1}{*}{14} & 6.2E-6 & 1.5E-4 & 1.7E-3 & 1.1E-2 & 4.6E-2 & 1.3E-1 & 2.3E-1 & 2.4E-1 & 1.4E-1 & 2.5E-2 \\
 & (5380) & (5924) & (6577) & (7376) & (8375) & (9661) & (11376) & (13776) & (17369) & (23337) \\
 \noalign{\smallskip}
  
\multirow{1}{*}{15} & 1.5E-6 & 4.1E-5 & 5.0E-4 & 3.6E-3 & 1.7E-2 & 5.5E-2 & 1.2E-1 & 1.7E-1 & 1.4E-1 & 6.1E-2 \\
 & (5069) & (5549) & (6118) & (6803) & (7645) & (8702) & (10070) & (11905) & (14497) & (18431) \\
 \noalign{\smallskip}
  
\multirow{1}{*}{16} & 3.8E-7 & 1.1E-5 & 1.5E-4 & 1.2E-3 & 6.4E-3 & 2.3E-2 & 6.0E-2 & 1.1E-1 & 1.2E-1 & 8.3E-2 \\
 & (4798) & (5225) & (5727) & (6323) & (7044) & (7932) & (9052) & (10509) & (12478) & (15286) \\
 \noalign{\smallskip}
  
\multirow{1}{*}{17} & 9.2E-8 & 3.1E-6 & 4.5E-5 & 3.9E-4 & 2.3E-3 & 9.4E-3 & 2.8E-2 & 5.9E-2 & 8.6E-2 & 8.1E-2 \\
 & (4559) & (4943) & (5390) & (5915) & (6541) & (7300) & (8238) & (9427) & (10982) & (13100) \\
 \noalign{\smallskip}
  
\multirow{1}{*}{18} & 2.0E-8 & 7.8E-7 & 1.3E-5 & 1.2E-4 & 7.9E-4 & 3.6E-3 & 1.2E-2 & 2.9E-2 & 5.1E-2 & 6.3E-2 \\
 & (4347) & (4695) & (5096) & (5564) & (6114) & (6772) & (7572) & (8565) & (9830) & (11493) \\
 \noalign{\smallskip}
  
\multirow{1}{*}{19} & 3.9E-9 & 1.8E-7 & 3.4E-6 & 3.7E-5 & 2.6E-4 & 1.3E-3 & 4.9E-3 & 1.3E-2 & 2.8E-2 & 4.1E-2 \\
 & (4159) & (4476) & (4839) & (5259) & (5748) & (6325) & (7018) & (7863) & (8916) & (10263) \\
 \noalign{\smallskip}
  
\multirow{1}{*}{20} & 5.5E-10 & 3.8E-8 & 8.7E-7 & 1.1E-5 & 8.3E-5 & 4.6E-4 & 1.9E-3 & 5.8E-3 & 1.4E-2 & 2.4E-2 \\
 & (3990) & (4281) & (4612) & (4991) & (5430) & (5943) & (6550) & (7280) & (8174) & (9292) \\
  
\hline
\end{tabular}
\\
{\small Values in brackets show the band origin in \AA.} \\
{\small Calculations are made using SEE solar flux on 23 June 2009 and at SZA = 60\dgr.}
\label{tab:WB-oi}
\end{table*}
\end{center}

\addtocounter{table}{-1}

\begin{center}
\begin{table*}
\caption{contd.} \vspace{1pt}
\small
\begin{tabular*}{\textwidth}{@{\extracolsep{\fill}}lccccccccccccccccccccc}
\hline \noalign{\smallskip}
$ \nu'\setminus\nu'' $ & 10 & 11 & 12 & 13 & 14 & 15 & 16 & 17 \\
\hline

\multirow{1}{*}{12} & 1.4E-4 & -- & -- & -- & -- & -- & -- & --   \\
 & (204814) & -- & -- & -- & -- & -- & -- & --  \\
\noalign{\smallskip}

\multirow{1}{*}{13} & 1.5E-3 & 1.5E-5 & -- & -- & -- & -- & -- & --  \\
 & (59621) & (378656) & -- & -- & -- & -- & -- & --  \\
\noalign{\smallskip}

\multirow{1}{*}{14} & 1.7E-5 & 1.1E-3 & 4.9E-8 &  -- & -- & -- & -- & --  \\
 & (35186) & (69984) & (2189094) & -- & -- & -- & -- & -- \\
\noalign{\smallskip}

\multirow{1}{*}{15} & 6.0E-3 & 5.4E-4 & 5.2E-4 &  -- & -- & -- & -- & --  \\
 & (25109) & (38919) & (84300) & -- & -- & -- & -- & -- \\
\noalign{\smallskip}

\multirow{1}{*}{16} & 2.5E-2 & 8.6E-4 & 7.9E-4 & 2.0E-4 & -- & -- & -- & --  \\
 & (19612) & (27132) & (43431) &  (105325) & -- & -- & -- & --  \\
\noalign{\smallskip}

\multirow{1}{*}{17} & 4.3E-2 & 8.8E-3 & 2.4E-6 & 6.6E-4 & 5.5E-5 & -- & -- & --  \\
 & (16153) & (20931) & (29461) &  (48989) & (139141) & -- & -- & -- \\
\noalign{\smallskip}

\multirow{1}{*}{18} & 4.9E-2 & 2.0E-2 & 2.4E-3 & 1.4E-4 & 4.1E-4 & 1.0E-5 & -- & --  \\
 & (13778) & (17109) & (22413) &  (32169) & (55990) & (202357) & -- & --  \\
\noalign{\smallskip}

\multirow{1}{*}{19} & 4.2E-2 & 2.7E-2 & 8.3E-3 & 4.0E-4 & 2.8E-4 & 2.1E-4 & 9.5E-7 & --  \\
 & (12047) & (14519) & (18167) &  (24089) & (35352) & (65067) & (362190) & --  \\
\noalign{\smallskip}

\multirow{1}{*}{20} & 3.0E-2 & 2.6E-2 & 1.3E-2 & 2.9E-3 & 9.9E-6 & 2.7E-4 & 8.6E-5 & 5.7E-9  \\
 & (10731) & (12649) & (15332) &  (19345) & (25997) & (39142) & (77278) & (1546312)  \\
\hline
\end{tabular*}
\end{table*}
\end{center}

\begin{center}
\begin{table*}
\caption{Overhead intensities (in R) of \nt\ $ B'^3\Sigma_u^{-} \rightarrow B^3\Pi_g $ band.} \vspace{1pt}
\small
\begin{tabular*}{\textwidth}{@{\extracolsep{\fill}}lccccccccccccccccccccc}
\hline \noalign{\smallskip}
$ \nu'\setminus\nu'' $ & 0 & 1 & 2 & 3 & 4 & 5 & 6 & 7 & 8 & 9 \\
\hline \noalign{\smallskip}
\multirow{1}{*}{0} & 1.2E-2 & 3.4E-3 & 2.5E-4 & 3.4E-6 & -- & -- & -- & -- & -- & --  \\
& (15280) & (20664) & (31616) & (65977) & -- & -- & -- & -- & -- & -- \\
 \noalign{\smallskip}

\multirow{1}{*}{1} & 6.2E-2 & 2.9E-3 & 1.1E-2 & 1.7E-3 & 3.1E-5 & -- & -- & -- & -- & -- \\
 & (12442) & (15793) & (21479) & (33241) & (71941) & -- & -- & -- & -- & -- \\
 \noalign{\smallskip}

\multirow{1}{*}{2} & 1.0E-1 & 8.6E-2 & 3.2E-3 & 1.3E-2 & 4.1E-3 & 1.0E-4 & -- &  -- & -- & -- \\
 & (10520) & (12820) & (16329) & (22337) & (34981) & (78778) & -- & -- & -- & -- \\
 \noalign{\smallskip}

\multirow{1}{*}{3} & 9.4E-2 & 2.2E-1 & 4.7E-2 & 2.5E-2 & 7.5E-3 & 6.1E-3 & 1.9E-4 & -- & -- & -- \\
 & (9132) & (10816) & (13212) & (16887) & (23237) & (36843) & (86653) & -- & -- & -- \\
 \noalign{\smallskip}
  
\multirow{1}{*}{4} & 5.9E-2 & 2.6E-1 & 2.4E-1 & 5.9E-3 & 4.8E-2 & 1.7E-3 & 6.4E-3 & 2.4E-4 & -- & -- \\
 & (8083) & (9375) & (11123) & (13618) & (17468) & (24181) & (38831) & (95765) &  -- & -- \\
 \noalign{\smallskip}
  
\multirow{1}{*}{5} & 2.9E-2 & 2.0E-1 & 3.8E-1 & 1.6E-1 & 4.0E-3 & 5.2E-2 & 3.0E-5 & 5.3E-3 & 2.5E-4 & -- \\
 & (7262) & (8289) & (9626) & (11440) & (14040) & (18072) & (25169) & (40949) & (106355) & -- \\
 \noalign{\smallskip}
  
\multirow{1}{*}{6} & 1.2E-2 & 1.1E-1 & 3.3E-1 & 3.6E-1 & 6.4E-2 & 2.8E-2 & 3.9E-2 & 1.8E-3 & 3.6E-3 & 2.1E-4 \\
 & (6603) & (7441) & (8501) & (9886) & (11769) & (14476) & (18699) & (26201) & (43199) & (118707) \\
 \noalign{\smallskip}
  
\multirow{1}{*}{7} & 4.3E-3 & 5.0E-2 & 2.2E-1 & 4.0E-1 & 2.6E-1 & 1.0E-2 & 4.8E-2 & 2.2E-2 & 4.4E-3 & 2.0E-3 \\
& (6062) & (6761) & (7625) & (8721) & (10154) & (12107) & (14927) & (19349) & (27275) & (45581) \\
 \noalign{\smallskip}
  
\multirow{1}{*}{8} & 1.4E-3 & 2.0E-2 & 1.1E-1 & 2.9E-1 & 3.6E-1 & 1.4E-1 & 5.4E-4 & 5.2E-2 & 8.8E-3 & 5.7E-3 \\
 & (5610) & (6204) & (6924) & (7815) & (8947) & (10430) & (12457) & (15392) & (20021) & (28389) \\
 \noalign{\smallskip}
  
\multirow{1}{*}{9} & 4.2E-4 & 7.0E-3 & 4.8E-2 & 1.7E-1 & 3.1E-1 & 2.6E-1 & 5.1E-2 & 1.1E-2 & 4.3E-2 & 2.2E-3 \\
 & (5227) & (5739) & (6350) & (7092) & (8011) & (9180) & (10714) & (12817) & (15872) & (20712) \\
 \noalign{\smallskip}
  
\multirow{1}{*}{10} & 1.2E-4 & 2.3E-3 & 1.8E-2 & 8.1E-2 & 2.0E-1 & 2.7E-1 & 1.6E-1 & 1.1E-2 & 2.3E-2 & 2.8E-2 \\
 & (4890) & (5345) & (5872) & (4500) & (7264) & (8213) & (9419) & (11007) & (13187) & (16364) \\
 \noalign{\smallskip}
  
\multirow{1}{*}{11} & 3.4E-5 & 7.1E-4 & 6.6E-3 & 3.4E-2 & 1.1E-1 & 2.0E-1 & 2.0E-1 & 7.6E-2 & 1.3E-4 & 2.6E-2 \\
 & (4614) & (5008) & (5467) & (6008) & (6655) & (7442) & (8420) & (9666) & (11308) & (13566) \\
 \noalign{\smallskip}
  
\multirow{1}{*}{12} & 9.2E-6 & 2.1E-4 & 2.2E-3 & 1.3E-2 & 4.9E-2 & 1.2E-1 & 1.7E-1 & 1.3E-1 & 2.9E-2 & 2.4E-3 \\
 & (4365) & (4716) & (5121) & (5592) & (6149) & (6815) & (7625) & (8633) & (9919) & (11616) \\
 \noalign{\smallskip}
  
\multirow{1}{*}{13} & 2.5E-6 & 6.2E-5 & 7.0E-4 & 4.7E-3 & 2.0E-2 & 5.8E-2 & 1.1E-1 & 1.2E-1 & 6.9E-2 & 7.2E-3 \\
 & (4145) & (4460) & (4821) & (5236) & (5721) & (6294) & (6979) & (7817) & (8853) & (10179) \\
 \noalign{\smallskip}
  
\multirow{1}{*}{14} & 6.5E-7 & 1.8E-5 & 2.2E-4 & 1.6E-3 & 7.9E-3 & 2.6E-2 & 6.0E-2 & 9.1E-2 & 8.2E-2 & 3.2E-2 \\
 & (3950) & (4235) & (4559) & (4929) & (5356) & (5854) & (6443) & (7148) & (8007) & (9078) \\
 \noalign{\smallskip}
  
\multirow{1}{*}{15} & 1.7E-7 & 4.9E-6 & 6.5E-5 & 5.3E-4 & 2.9E-3 & 1.1E-2 & 3.0E-2 & 5.5E-2 & 6.8E-2 & 4.8E-2 \\
 & (3775) & (4035) & (4328) & (4660) & (5040) & (5479) & (5991) & (6596) & (7322) & (8206) \\
 \noalign{\smallskip}
  
\multirow{1}{*}{16} & 4.4E-8 & 1.4E-6 & 1.9E-5 & 1.7E-4 & 1.0E-3 & 4.3E-3 & 1.3E-2 & 2.9E-2 & 4.5E-2 & 4.6E-2 \\
 & (3619) & (3857) & (4123) & (4424) & (4765) & (5155) & (5606) & (6132) & (6754) & (7500) \\
 \noalign{\smallskip}
  
\multirow{1}{*}{17} & 1.1E-8 & 3.7E-7 & 5.6E-6 & 5.2E-5 & 3.4E-4 & 1.6E-3 & 5.4E-3 & 1.4E-2 & 2.6E-2 & 3.4E-2 \\
 & (3477) & (3696) & (3940) & (4214) & (4522) & (4872) & (5273) & (5737) & (6277) & (6917) \\
 \noalign{\smallskip}
  
\multirow{1}{*}{18} & 2.8E-9 & 9.9E-8 & 1.6E-6 & 1.6E-5 & 1.1E-4 & 5.6E-4 & 2.1E-3 & 6.2E-3 & 1.3E-2 & 2.1E-2 \\
 & (3349) & (3552) & (3776) & (4027) & (4308) & (4624) & (4984) & (5395) & (5871) & (6427) \\
 \noalign{\smallskip}
  
\multirow{1}{*}{19} & 6.6E-10 & 2.6E-8 & 4.5E-7 & 4.8E-6 & 3.6E-5 & 2.0E-4 & 8.1E-4 & 2.6E-3 & 6.3E-3 & 1.2E-2 \\
 & (3232) & (3420) & (3629) & (3859) & (4116) & (4404) & (4729) & (5099) & (5521) & (6010) \\
 \noalign{\smallskip}
  
\multirow{1}{*}{20} & 1.4E-10 & 6.4E-9 & 1.2E-7 & 1.4E-6 & 1.1E-5 & 6.6E-5 & 3.0E-4 & 1.0E-3 & 2.8E-3 & 5.9E-3 \\
 & (3125) & (3301) & (3495) & (3708) & (3945) & (4209) & (4504) & (4838) & (5217) & (5651) \\
  
\hline
\end{tabular*}
\\
{\small Values in brackets show the band origin in \AA.} \\
{\small Calculations are made using SEE solar flux on 23 June 2009 and at SZA = 60\dgr.}
\label{tab:BB-oi}
\end{table*}
\end{center}

\addtocounter{table}{-1}

\begin{center}
\begin{table*}
\caption{contd.} \vspace{1pt}
\small
\begin{tabular}{lccccccccccccccccccccc}
\hline \noalign{\smallskip}
$ \nu'\setminus\nu'' $ & 10 & 11 & 12 & 13 & 14 & 15 &
16 & 17 & 18 & 19  & 20 \\
\hline

\multirow{1}{*}{7} & 1.5E-4 & -- & -- &  -- & -- & -- & -- & -- & -- & -- &  -- \\
 & (133152) & -- & -- &  -- & -- & -- & -- & -- & -- & -- &  -- \\
\noalign{\smallskip}

\multirow{1}{*}{8} & 9.6E-4 & 9.9E-5 & -- &  -- & -- & -- & -- & -- & -- & -- &  -- \\
 & (48087) & (150065) & -- & -- &  -- & -- & -- & -- & -- & -- & -- \\
\noalign{\smallskip}

\multirow{1}{*}{9} & 5.7E-3 & 3.8E-4 & 5.9E-5 & -- &  -- & -- & -- & -- & -- & -- & -- \\
 & (29540) & (50709) & (169831) &  -- &  -- & -- & -- & -- & -- & -- & -- \\
\noalign{\smallskip}

\multirow{1}{*}{10} & 1.1E-4 & 4.6E-3 & 1.2E-4 & 3.2E-5 & -- & -- &  -- & -- & -- & -- & --  \\
 & (21422) & (30723) & (53427) &  (192800) & -- & -- &  -- & -- & -- & -- & --  \\
\noalign{\smallskip}

\multirow{1}{*}{11} &1.6E-2 & 1.9E-4 & 3.3E-3 & 2.5E-5 & 1.7E-5 & -- &  -- & -- & -- & -- & --  \\
 & (16868) & (22148) & (31929) &  (56215) & (219162) & -- &  -- & -- & -- & -- & --  \\
\noalign{\smallskip}

\multirow{1}{*}{12} & 2.3E-2 & 7.3E-3 & 7.6E-4 & 2.1E-3 & 1.4E-6 & 8.5E-6 &  -- & -- & -- & -- & --  \\
 & (13955) & (17383) & (22886) &  (33152) & (59037) & (248754) & -- &  -- & -- & -- & -- \\
\noalign{\smallskip}

\multirow{1}{*}{13} & 6.6E-3 & 1.7E-2 & 2.8E-3 & 1.1E-3 & 1.3E-3 & 9.9E-7 & 4.2E-6 & -- &  -- & -- & -- \\
 & (11932) & (14352) & (17907) &  (23633) & (34378) & (61844) & (280746) & -- &  -- & -- & -- \\
\noalign{\smallskip}

\multirow{1}{*}{14} & 6.0E-4 & 8.5E-3 & 1.0E-2 & 7.9E-4 & 1.1E-3 & 7.2E-4 & 4.4E-6 & 2.1E-6 & -- &  -- & -- \\
 & (10446) & (12255) & (14756) &  (18437) & (24383) & (35595) & (64575) & (313255) & -- &  -- & -- \\
\noalign{\smallskip}

\multirow{1}{*}{15} & 1.3E-2 & 2.2E-4 & 8.1E-3 & 5.9E-3 & 1.2E-4 & 9.8E-4 & 3.9E-4 & 6.2E-6 & 1.1E-6 &-- &  -- \\
 & (9308) & (10718) & (12584) &  (15166) & (18972) & (25131) & (36786) & (67157) & (343066) & -- &  -- \\
\noalign{\smallskip}

\multirow{1}{*}{16} & 2.5E-2 & 3.6E-3 & 1.3E-3 & 6.3E-3 & 3.0E-3 & 1.2E-0 & 7.5E-4 & 2.0E-4 & 6.2E-6 & 6.4E-7 & -- \\
 & (8410) & (9545) & (10996) &  (12918) & (15581) & (19507) & (25869) & (37932) & (69506) & (365813) &  --  \\ 
\noalign{\smallskip}

\multirow{1}{*}{17} & 2.8E-2 & 1.2E-2 & 5.6E-4 & 2.1E-3 & 4.4E-3 & 1.3E-3 & 4.0E-5 & 5.2E-4 & 1.0E-4 & 5.1E-6 & 4.0E-7 \\
 & (7683) & (8619) & (9786) &  (11280) & (13258) & (15997) & (20039) & (26589) & (39011) & (71531) &  (376997)  \\
\noalign{\smallskip}

\multirow{1}{*}{18} & 2.4E-2 & 1.6E-2 & 4.7E-3 & 1.3E-9 & 2.2E-3 & 2.7E-3 & 5.3E-4 & 8.8E-5 & 3.4E-4 & 5.0E-5 & 3.8E-6 \\
 & (7084) & (7872) & (8834) &  (10033) & (11568) & (13600) & (16414) & (20564) & (27283) & (40001) &  (73141)  \\ 
\noalign{\smallskip}

\multirow{1}{*}{19} & 1.6E-2 & 1.5E-2 & 8.2E-3 & 1.5E-3 & 1.9E-4 & 1.9E-3 & 1.6E-3 & 1.8E-4 & 1.1E-4 & 2.2E-4 & 2.4E-5 \\
 & (6581) & (7255) & (8065) &  (9052) & (10284) & (11859) & (13944) & (16829) & (21077) & (27939) &  (40875)  \\
\noalign{\smallskip}

\multirow{1}{*}{20} & 9.6E-3 & 1.1E-2 & 8.9E-3 & 3.8E-3 & 3.2E-4 & 4.2E-4 & 1.4E-3 & 8.4E-4 & 4.7E-5 & 1.0E-4 & 1.3E-4 \\
 & (6153) & (6739) & (7432) &  (8262) & (9276) & (10539) & (12154) & (14289) & (17238) & (21572) &  (28548)   \\ 
\hline
\end{tabular}
\\
{\small Values in brackets show the band origin in \AA.} \\
{\small Calculations are made using SEE solar flux on 23 June 2009 and at SZA = 60\dgr.}
\end{table*}
\end{center}

\begin{center}
\begin{table*}
\caption{Overhead intensities (in R) of $ E^3\Sigma_g^+\rightarrow A^3\Sigma_u^+$,
$ E^3\Sigma_g^+\rightarrow B^3\Pi_g$, and $ E^3\Sigma_g^+\rightarrow C^3\Pi_u$ bands of \nt.} \vspace{1pt}
\small
\begin{tabular*}{\textwidth}{@{\extracolsep{\fill}}lccccccccccccccccccccc}
\hline \noalign{\smallskip}
$ \nu'\setminus\nu'' $ & 0 & 1 & 2 & 3 & 4 & 5 & 6 & 7 & 8 & 9 & 10 \\
\hline \noalign{\smallskip}
\multicolumn{12}{c}{$ E^3\Sigma_g^+\rightarrow A^3\Sigma_u^+$} \\
\multirow{1}{*}{0} & 8.9E-3 & 2.8E-2 & 4.5E-2 & 5.0E-2 & 4.1E-2 & 2.8E-2 & 1.6E-2 & 7.7E-3 & 3.3E-3 & 1.2E-3 & 3.9E-4 \\
& (2173) & (2243) & (2316) & (2392) & (2472) & (2555) & (2643) & (2734) & (2830) & (2930) & (3035) \\
 \noalign{\smallskip}

\multirow{1}{*}{1} & 2.7E-3 & 6.2E-3 & 7.0E-3 & 4.8E-3 & 2.2E-3 & 5.6E-4 & 3.9E-5 & 2.0E-5 & 8.6E-5 & 1.0E-4 & 7.8E-5 \\
& (2074) & (2138) & (2204) & (2273) & (2345) & (2420) & (2498) & (2580) & (2665) & (2754) & (2846) \\
 \noalign{\smallskip}

\multicolumn{12}{c}{$ E^3\Sigma_g^+\rightarrow B^3\Pi_g$} \\

\multirow{1}{*}{0} & 7.5E-3 & 1.1E-2 & 9.2E-3 & 5.7E-3 & 2.9E-3 & 1.3E-3 & 5.6E-4 & 2.2E-4 & 8.3E-5 & 3.0E-5 & 1.0E-5 \\
& (2742) & (2877) & (3022) & (3181) & (3353) & (3542) & (3749) & (3978) & (4230) & (4511) & (4826) \\
 \noalign{\smallskip}

\multirow{1}{*}{1} & 5.6E-4 & 2.0E-4 & 1.0E-6 & 5.8E-5 & 1.2E-4 & 1.2E-4 & 8.0E-5 & 4.6E-5 & 2.3E-5 & 1.1E-5 & 4.6E-6 \\
& (2587) & (2707) & (2835) & (2974) & (3124) & (3288) & (3465) & (3660) & (3872) & (4107) & (4365) \\
 \noalign{\smallskip}
 
\multicolumn{12}{c}{$ E^3\Sigma_g^+\rightarrow C^3\Pi_u$} \\
\multirow{1}{*}{0} & 1.3E-1 & 1.1E-2 & 4.1E-4 & 2.6E-6 & -- & -- & -- & -- & -- & -- & -- \\
& (14713) & (20824) & (34947) & (101275) & -- & -- & -- & -- & -- & -- & --   \\
 \noalign{\smallskip}
  
\multirow{1}{*}{1} & 2.3E-3 & 2.3E-3 & 5.5E-4 & 3.9E-5 & 7.4E-7 & -- & -- & -- & -- & -- & --  \\
& (11134) & (14312) & (19816) & (31522) & (71884) & -- & -- & -- & -- & -- & --  \\
 \noalign{\smallskip}

\hline
\end{tabular*}
\\
{\small Values in brackets show the band origin in \AA.} \\
{\small Calculations are made using SEE solar flux on 23 June 2009 and at SZA = 60\dgr.}
\label{tab:EABC-oi}
\end{table*}
\end{center}

\begin{center}
\begin{table*}
\caption{Overhead intensities (in R) of \nt\ Reverse First Positive ($ A^3\Sigma_u^+ 
\rightarrow B^3\Pi_g $) band.} \vspace{1pt}
\small
\begin{tabular*}{\textwidth}{@{\extracolsep{\fill}}lccccccccccccccccccccc}
\hline \noalign{\smallskip}
$ \nu'\setminus\nu'' $ & 0 & 1 & 2 & 3 & 4 & 5 & 6 & 7 & 8  \\
\hline \noalign{\smallskip}
\multirow{1}{*}{8} & 1.5E+0 & -- & -- & -- & -- & -- & -- & -- & -- \\
& (88467) & -- & -- & -- & -- & -- & -- & -- & -- \\
 \noalign{\smallskip}

\multirow{1}{*}{9} & 1.9E+0 & 7.0E-1 & -- & -- & -- & -- & -- & -- & -- \\
 & (42771) & (158042) & -- & -- & -- & -- & -- & -- & -- \\
 \noalign{\smallskip}

\multirow{1}{*}{10} & 8.2E-1 & 2.3E+0 & 8.5E-3 & -- & -- & -- & -- & -- & --\\
 & (28438) & (55215) & (741757) & -- & -- & -- & -- & -- & -- \\
 \noalign{\smallskip}

\multirow{1}{*}{11} & 2.5E-1 & 1.5E+0 & 1.2E+0 & -- & -- & -- & -- & -- & -- \\
 & (21436) & (33788) & (77920) & -- & -- & -- & -- & -- & -- \\
 \noalign{\smallskip}
  
\multirow{1}{*}{12} & 7.7E-2 & 6.9E-1 & 1.6E+0 & 3.1E-1 & -- & -- & -- & -- & -- \\
 & (17292) & (24523) & (41640) & (132578) & -- & -- & -- & -- & -- \\
 \noalign{\smallskip}
  
\multirow{1}{*}{13} & 2.2E-2 & 2.5E-1 & 9.5E-1 & 9.9E-1 & 7.9E-3 & -- & -- & -- & --\\
 & (14555) & (19361) & (28664) & (54305) & (448147) & -- & -- & -- & -- \\
 \noalign{\smallskip}
  
\multirow{1}{*}{14} & 6.3E-3 & 8.7E-2 & 4.5E-1 & 9.0E-1 & 3.9E-1 & -- & -- & -- & --  \\
& (12617) & (16076) & (22007) & (34521) & (78216) & -- & -- & -- & --  \\
 \noalign{\smallskip}
  
\multirow{1}{*}{15} & 1.9E-3 & 3.0E-2 & 1.9E-1 & 5.5E-1 & 6.3E-1 & 7.1E-2 & -- & -- & --  \\
& (11175) & (13806) & (17964) & (25513) & (43455) & (140462) & -- & -- & --  \\
 \noalign{\smallskip}
  
\multirow{1}{*}{16} & 5.7E-4 & 9.9E-3 & 7.3E-2 & 2.8E-1 & 5.1E-1 & 3.0E-1 & 5.0E-4 &-- & --  \\
& (10063) & (12148) & (15254) & (20373) & (30393) & (58790) & (710500) & -- & --  \\
 \noalign{\smallskip}
  
\multirow{1}{*}{17} & 1.8E-4 & 3.4E-3 & 2.8E-2 & 1.3E-1 & 3.1E-1 & 3.6E-1 & 8.3E-2 & -- & --  \\
& (9181) & (10886) & (13316) & (17057) & (23560) & (37662) & (91328) & -- & --  \\
 \noalign{\smallskip}
  
\multirow{1}{*}{18} & 5.8E-5 & 1.2E-3 & 1.1E-2 & 5.5E-2 & 1.7E-1 & 2.8E-1 & 1.9E-1 & 6.8E-3 & --  \\
& (8468) & (9897) & (11865) & (14747) & (19370) & (27985) & (49675) & (207470) & --  \\
 \noalign{\smallskip}
  
\multirow{1}{*}{19} & 1.9E-5 & 4.1E-4 & 4.0E-3 & 2.3E-2 & 8.3E-2 & 1.8E-1 & 2.0E-1 & 6.1E-2 & --  \\
& (7880) & (9103) & (10763) & (13052) & (16548) & (22452) & (34558) & (73390) & --  \\
 \noalign{\smallskip}
  
\multirow{1}{*}{20} & 6.5E-6 & 1.5E-4 & 1.5E-3 & 9.8E-3 & 4.0E-2 & 1.0E-1 & 1.6E-1 & 1.0E-1 & 8.3E-3 \\
& (7390) & (8456) & (9852) & (11761) & (14525) & (18884) & (26773) & (45371) & (142399) \\
 \noalign{\smallskip}
    
\hline
\end{tabular*}
\\
{\small Values in brackets show the band origin in \AA.} \\
{\small Calculations are made using SEE solar flux on 23 June 2009 and at SZA = 60\dgr.}
\label{tab:r1p-oi}
\end{table*}
\end{center}

\clearpage
\newpage

\end{document}